\documentclass[useAMS,usenatbib]{mn2e}
\usepackage[dvipdfm]{graphicx}

\usepackage[greek,english]{babel}
\usepackage{longtable}
\usepackage{rotating}
\usepackage{amssymb}
\usepackage{lscape}
\usepackage{amsmath}

 \def\mic{\,\mbox{\textgreek{m}m}}

\newcommand{\degree}{\ensuremath{^\circ}}

\title[Multiwavelength analysis of the nebula around G79.29+0.46]{The candidate luminous blue variable G79.29+0.46: a comprehensive study of its ejecta through a multiwavelength analysis}
\author[C. Agliozzo et al.]{C.~Agliozzo,$^{1,2}$\thanks{E-mail:
c.agliozzo@gmail.com} A.~Noriega-Crespo,$^{3}$ 
G.~Umana,$^4$ N.~Flagey,$^{5}$ C.~Buemi,$^{4}$ 
\newauthor
A.~Ingallinera,$^{4}$ 
 C.~Trigilio,$^{4}$  P.~Leto$^{4}$\\
$^{1}$Departamento de Ciencias Fisicas, Universidad Andres Bello, Avda. Republica 252, Santiago 8320000, Chile\\
$^{2}$Dipartimento di Fisica e Astronomia, Sezione Astrofisica, Universit\`{a}
degli studi di Catania, via S. Sofia 64, 95123, Catania, Italy\\
$^{3}$Infrared Processing and Analysis Center, California Institute of Technology, 1200, East California Boulevard, Pasadena, CA 91125, USA\\
$^{4}$INAF Osservatorio Astrofisico di Catania, via S. Sofia 78, 95123, Catania,
Italy\\
$^{5}$Jet Propulsion Laboratory, California Institute of Technology, 4800 Oak Grove Drive, Pasadena, CA 91109, USA\\
}

\begin{document}

\date{}

\pagerange{\pageref{firstpage}--\pageref{lastpage}} \pubyear{2014}

\maketitle

\label{firstpage}

\begin{abstract}
We present a multiwavelength analysis of the nebula around
the candidate luminous blue variable G79.29+0.46. The
study is based on our radio observations performed at the
Expanded Very Large Array and at the Green Bank Telescope and on archival infrared datasets, including recent images obtained by the \emph{Herschel} Space Observatory.

We confirm that the radio central object is characterized by a stellar wind and derive a current mass-loss rate of about $1.4\times10^{-6}\, \rm M_{\odot} yr^{-1}$. We find the presence of a dusty compact envelope close to the star, with a temperature between $\sim$ 40 and 1200 K.  We estimate for the outer ejecta an ionised gas mass of 1.51 $\rm M_{\odot}$ and a warm (60--85 K) dust mass of 0.02 $\rm M_{\odot}$. Diagnostics of the far-infrared spectra indicate the presence of a photo-dissociation region around the ionised gas.

Finally, we model the nebula with the photo-ionization code CLOUDY, using as input parameters those estimated from
our analysis. We find for the central star a luminosity of 10$^{5.4} \rm L_{\odot}$ and an effective temperature of 20.4 kK.
\end{abstract}


\section{Introduction}

Evolved luminous stars with an initial mass more than 20 $\rm M_{\odot}$ are believed to enter an instability phase, historically known as luminous blue variable (LBV).  This phase is characterized by intense mass-loss rates ($\sim10^{-5}-10^{-3}\, \rm M_{\odot} yr^{-1}$), associated to visual/near-IR spectroscopic and photometric variabilities \citep{1994HD}, before the star evolves as Wolf-Rayet (WR), once it has reduced its mass \citep{1976Conti}. However, the occurrence of this phase during the massive stars evolution and the mass-loss mechanism are still not well understood, mostly because of the rarity of LBVs in our Galaxy.  
For the moment, the objects that satisfy the variability criteria coupled with high mass-loss rates are only 12 in the Milky Way,   but recently the number of candidate LBVs (cLBVs) has been increased to 55 \citep{1994HD, vanGenderen2001, Clark2005, 2010Gvaramadze, 2011Wachter, 2012Naze}, based on the discovery of ring nebulae surrounding these luminous stars. In fact, as a result
of the mass-loss, such evolved massive stars can be surrounded
by circumstellar nebulae, rich of processed material \citep{2001Lamers}. LBV nebulae (LBVNe) have been observed in about 20 objects among LBVs and cLBVs in our Galaxy and they have shown to be dusty, hence emitting in a wide spectral domain. Such nebulae are indeed the fingerprints of the mass-loss phenomenon, therefore multiwavelength observations provide unique insights in the LBV mass-loss history.

The importance of studying LBVNe is increased by the possibility that they can skip the WR stage and experience the explosive supernova (SN) event \citep[e.g. ][]{2006KotakVink,2008Vinkbis, 2013vanDyk}. If this is true, the LBV mass-loss phenomenon is of great interest also for the study of type IIn SNe, whose evolution is influenced by the circumstellar material originated through mass-loss of the progenitor star \citep{2003Chugai, 2007SmithN, 2013Ofek}. The mass-loss archaeology appears to be, therefore, fundamental for different aspects of the massive stars evolution.

In this work we propose a new study of the ejecta around G79.29+0.46, a galactic candidate LBV located in the Cygnus-X \citep[$D\sim1.7$
kpc,][]{2006Schn}, one of the richest star forming regions in our Galaxy. Thanks
to its location and also to its beautiful aspect and intriguing nature, this
star and its associated nebula have been widely observed in spectral
domains ranging from the optical to the radio. These observations have allowed to investigate different components of the ejecta associated to G79.29+0.46, which showed to have lost mass with different events \citep{2010Kraemer, 2010Esteban}. Diverse works focused on this source can be mentioned \citep[e.g.][]{1994Higgs, 1996Waters, 2000Voors, 2008Rizzo, 2011UmanaG79}, but a comprehensive study of the star and its associated nebula has never been performed as yet. Moreover, the physical parameters determined in these works are often based on different assumptions, making difficult their comparison and interpretation. 
For instance, \citet{1994Higgs} interpreted the ring-like shell around the star as interstellar material swept-up  by the stellar wind and estimated for this shell a mass of about 14-15 $\rm\,M_{\odot}$. Later, the nebula has been recognized as ejecta associated to the star \citep{1996Waters}, but a reappraisal of its properties is still missing. The nebular mass inferred by \citet{1994Higgs} and \citet{1996Waters} suggests that the nebula associated to G79.29+0.46 is amongst the most massive known \citep[cf. Table 9 in][]{2003Clark}. If correct, the mass would rival that of the Homunculus around $\eta$ Carina \citep[e.g.][]{1999Morris,2003Smithhomu}, but the central star of G79.29+0.46 is likely over an order of magnitude fainter \citep[about $10^{5.5}\rm\,L_{\odot}$, while $\eta$ Car is $10^{6.7}\rm\,L_{\odot}$,][]{1994Higgs,2000Voors, 2001Hillier} and consequently far less massive. Therefore, the nebular mass would represent a higher proportion of its initial mass.

In this paper we propose a new analysis of the multiple components of the nebula around G79.29+0.46, by means of recent observations which we performed in the radio domain, new far-IR images provided by the \emph{Herschel} Space Observatory and other archival far- and mid-IR spectroscopic and photometric datasets collected in the last 20 years by different telescopes. The main assumption in this work is that the photometric and spectroscopic properties of this star has not changed significantly in the considered time-lapse, hypothesis reinforced by the fact that G79.29+0.46 is still a candidate LBV, since no significant visual or near-IR variability has been observed so far \citep{2000Voors, 2008Vink}. 

This work started with our observations of G79.29+0.46 performed at the Expanded Very Large Array (EVLA). Preliminary results of these observations were  presented in \citet{2011UmanaG79} (Paper 1 hereafter), and the highest resolution radio image was compared with the \emph{Spitzer} infrared images, revealing us that the multiple shells around the star have signature of different populations of dust and that the radio emission is ionization bounded. In this paper we complement the preliminary work in Paper 1 with new radio observations performed at the Green Bank Telescope (GBT) to overcome the typical problem of interferometers, which is the ``zero baseline missing", resulting in the impossibility to determine the total flux of extended sources. 
Therefore we propose a new analysis of the radio emission aimed at determining important physical parameters related to the ionized gas and to the central star.  We also perform an analysis of the IR datasets retrieved from the data archives, in order to determine  IR lines and physical properties related to the dust component.

The paper is organized as in the following: in Section \ref{par:obs} we describe our observations and the archival  datasets considered in this study. Then we present our analysis of the radio datasets (Section \ref{par:maps}), focused on the ionized gas distribution and emission, while in Section \ref{par:infrared} we study the dust emission and properties, probed by the IR images. Section \ref{sec:irspectra} concerns the spectroscopic mid- and far-IR analysis of the nebula and in Section \ref{sec:sed} we present a simple model obtained with the photo-ionization code Cloudy. We conclude with a brief description of the star mass-loss history in Section \ref{sec:discus}.

\section{The datasets}
\label{par:obs}
\subsection{Radio observations and data processing}

\begin{table*}
\begin{center}
\caption{Properties of the interferometric radio maps. }
\label{tab:EVLAmaps}

\begin{tabular}{lccccccc}

\hline
   Configuration  & HPBW & PA  & LAS&Peak on-source & rms \\
   $\&$ frequency & (arcsec)$^{2}$ &   (deg)       &  (arcsec)   & (mJy
beam$^{-1}$)    &    (mJy beam$^{-1}$)                 \\
 \hline
 EVLA:D, 1.391 GHz  &  50.40$\times$42.83    &  163.6        &  129       &       
    26.0            & 2\\
 EVLA:D, 4.959 GHz  &  16.46$\times$14.16    &  $-33.2$           &  88   &           
6.8            & 0.7\\
 EVLA:C, 1.391 GHz  &  27.72$\times$14.44    &  44.4           &  258    &        
   9.6            & 2.5\\
 EVLA:C, 4.959 GHz  &  9.41$\times$4.84    &  58.7                & 124 &           
2.11            & 0.18\\
 EVLA:D+C, 1.391 GHz  &  26.84$\times$24.65    &  36.1           &  129    &      
     13.17            & 1.1\\
 EVLA:D+C, 4.959 GHz  &  4.56$\times$3.09    &  101.16              &  124 &        
   1.76          & 0.07\\
 VLA:D+C, 8.46 GHz  &  6.25$\times$3.12    &  57.3              &  164 &        
   1.68          & 0.15\\
     \hline\\
\end{tabular}

\end{center}

\end{table*}

\begin{table*}
\begin{center}
\caption{Properties of the GBT and GBT\&EVLA radio maps. }
\label{tab:EVLAGBTmaps}

\begin{tabular}{lccccccc}

\hline
   Configuration  & HPBW & PA  & FOV&Peak on-source & st dev \\
   $\&$ frequency & (arcsec)$^{2}$ &   (deg)       &  (arcsec)   & (mJy
beam$^{-1}$)    &    (mJy beam$^{-1}$)                 \\
 \hline

 GBT, 5.100 GHz  &  144$\times$144    &  0                &    1500      &
14.27$\times10^{3}$           & 12.43\\

 GBT\& EVLA:D+C, 4.959 GHz  & 4.56$\times$3.09    &  101.16             & 1500&     
      3.22          & 0.06\\
     \hline\\
\end{tabular}

\end{center}
\end{table*}

We performed high-resolution and high-sensitivity observations of G79.29+0.46 at 20 and 6 cm (respectively 1.4 and 5 GHz) using the EVLA in configurations D and C. These observations and the data reduction process were already described
in Paper 1, where we presented only the final map at 6 cm obtained by concatenating the uv-datasets in both the configurations D and C. Here we describe the imaging process for each dataset. 

The calibrated visibilities were imaged adopting a natural weighting scheme and using the \emph{Clark}
algorithm for the dirty image deconvolution. The cleaned images were therefore
convolved with a 2D-Gaussian, with the half power beam width (HPBW) reported in
Table \ref{tab:EVLAmaps}. Following this scheme, we have obtained four maps. 
For each frequency we also concatenated data obtained in the two array configurations D and C, in order to increase the \emph{uv}-coverage. Therefore, we imaged the new datasets. 
Peak flux density (on-source) and the rms-noise are listed in Table \ref{tab:EVLAmaps}
for all the maps. We also report the largest angular scale (LAS) achieved with the proposed array configurations. 

To complement the interferometric datasets, single-dish observations of G79.29+0.46 were carried out 
at the GBT at 1.4 and 5 GHz in June 2011 as a part of a project including a sample of galactic circumstellar shells (Ingallinera et al. 2014, in prep). 
The observations were performed running 30 scans on both RA and DEC directions for a final field of view  (FOV) of $1\degree 15'$ at 1.4 GHz and $25'$ at 5
GHz.  A data reduction, which consisted in a basic calibration and imaging, has been performed using the IDL routine \texttt{topsi} provided by the GBT staff. The images were however affected by ``stripes'', systematic errors due to instrumental drifts which shift the base-level of each scan. The images were therefore ``destriped'' using an appropriate filter in the Fourier domain (\citealt{1988Emerson}; Ingallinera et al. 2014, in prep.).  
Unfortunately, at 1.4 GHz the beam of the telescope is about four times bigger than the nebula size and as a result the sensitivity is limited by confusion. Therefore, we complement the interferometric map with the single-dish one only at 5 GHz, as in the following.

The 5 GHz GBT map was rescaled to the EVLA map units, i.e. converted from unit of antenna temperature $T_A$  to Jy beam$^{-1}$,  
by using the GBT antenna gain $G=2.84\,\eta_A$, with the antenna efficiency
$\eta_A=0.72$\footnote{See observing
manual for the GBT at www.gb.nrao.edu/gbtprops/man/GBTpg.pdf}, valid at 5 GHz. The GBT map was then re-scaled to the central frequency of the EVLA data
(the GBT central frequencies is 5.100 GHz and that EVLA 
is 4.959 GHz). This conversion was performed assuming that the
emission is almost due to optically thin free-free transitions ($\alpha=-0.1$), 
assumption confirmed in Section 3.2 by analysing the spectral index from the EVLA maps. The resulting flux density peak  in the 5 GHz GBT map is shown in Table \ref{tab:EVLAGBTmaps}. 
Since in the map the background emission is high, the rms does not give a good estimation of the noise, hence in Table \ref{tab:EVLAGBTmaps} we provide the standard deviation (st dev). 
The new single-dish map was then imported into the CASA package (version 4.2) and
processed together with the EVLA map at 5 GHz, by using the recently improved task
\texttt{feather}, which combines two images with different
spatial resolutions. In this procedure the single-dish flux scale was not modified (\texttt{sdfactor}=1) and the effective single-dish diameter was set to 100 m.

\begin{figure*}
\centering
\begin{minipage}{.4\linewidth}
  \includegraphics[width=\textwidth]{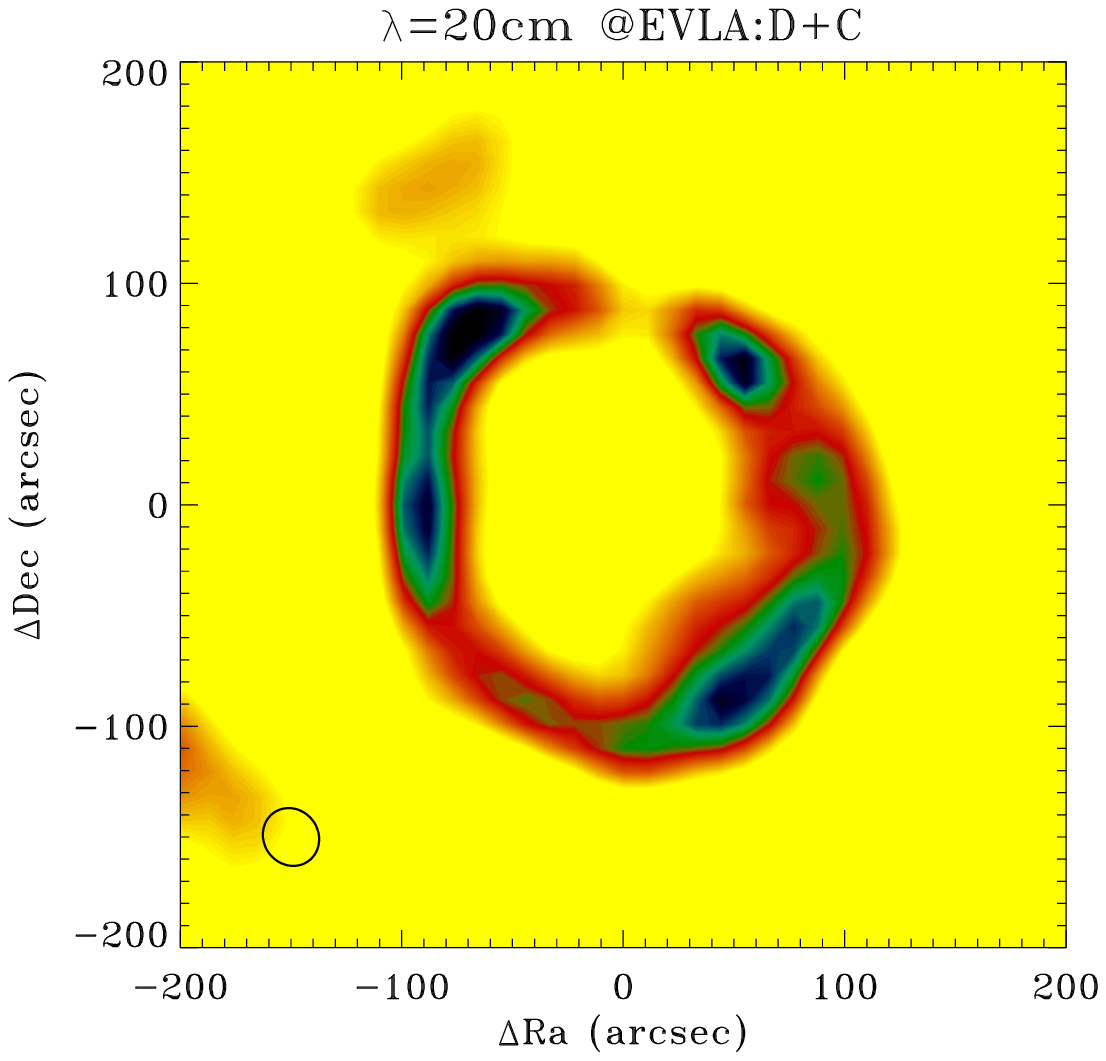}
\end{minipage}
\begin{minipage}{.4\linewidth}
 \includegraphics[width=\textwidth]{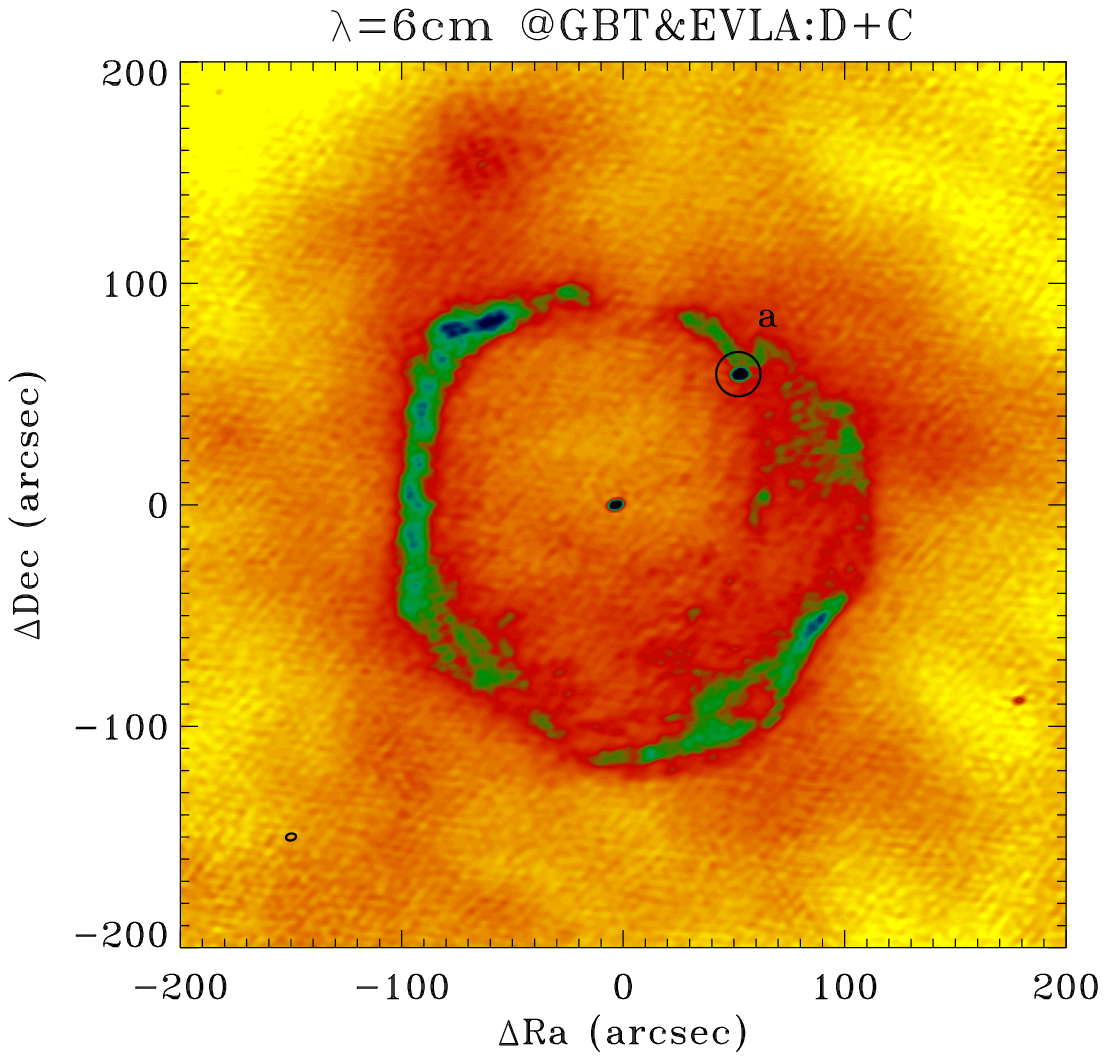}
\end{minipage}
\caption{Final radio maps obtained in this work. On the left: EVLA:D$\&$C 20 cm (1.4 GHz) map. On the right: 6 cm (5 GHz) map after combining GBT and EVLA images. In the bottom-left corner
of each map the synthetic beam is shown, i.e. the resolution in the map. In the right panel object ``a" is an extra-galactic object, first detected by \citet{1994Higgs}.}
 \label{fig:radiomaps}
\end{figure*}
             
The final map at 5 GHz is shown in Fig. \ref{fig:radiomaps}, together with that at 1.4 GHz. 
In the 5 GHz map the peak flux density corresponds to the point-source projected on the nebula in the north-west direction (object ``a"). 
The map is filled with diffuse emission arising from the close HII region in the south of G79.29+0.46 \citep[e.g. Fig. 1 in][]{1994Higgs}. As a result, there are not regions free of emission where evaluate properly the rms in the map. Therefore, we approximate again the noise as the standard deviation (Table \ref{tab:EVLAGBTmaps}). 
For the analysis of the radio emission (Section \ref{par:maps}), the background contribution will be subtracted from the map.

Exploring the VLA data archives, we found further observations of G79.29+0.46 at 8.46 GHz. These
observations were performed in May and August 2005, using configurations C
and B (P.I. Lang C.) in continuum mode. We have retrieved these datasets
and calibrated them adopting the same procedure used for the EVLA:D$\&$C data. 
Properties of the final map are summarized in Table \ref{tab:EVLAmaps}.

\subsection{Archival infrared datasets}
\begin{table*}
             
                  \begin{center}
     \caption{Archival spectroscopic data and instrument properties summary.}
                  \label{tab:spectraproperties}
                   \begin{tabular}{lcccr}
                      \hline
                      Telescope&Instrument&$\lambda$&FOV&Project\\
                        & &   ($\mic$)            & (arcsec)$^2$ &   ID    \\
                       \hline
                       \emph{Spitzer} & IRS/SL & 5.2--14.5& 3.6$\times$57&17333-248/504$^a$\\
                       \emph{Spitzer} & IRS/SH & 9.9--19.6& 4.7$\times$11.3&17333-248/504$^a$\\  
                       \emph{ISO} & SWS06     &  12.2--13.2  &27$\times$14& 69200901    \\  
                       \emph{ISO} & SWS06    &  17--27; 30--34 &33$\times$20&76802901    \\  
                       \emph{Spitzer} & IRS/LH& 18.7--37.2& 11.1$\times$22.3&17333-248/504$^a$ \\  
                       \emph{ISO} & SWS06     &  34--35  &33$\times$20& 51705105    \\  
                       \emph{ISO} & LWS02     &  41.14--178.75  &84$\times$84 & 358010-03/05/07/  \\  
                          &      &    & & 358010-08/11/14   \\  


                       \hline
                    \end{tabular}
                    
                    \end{center}
                                      \footnotesize
                    $a$ Respectively ON and OFF nebula observations.
                    \end{table*}

\begin{table}
\begin{center}
\caption{Archival photometric data and instrument properties summary.}
\label{tab:mapsproperties}

\begin{tabular}{llccr}
   \hline
   Telescope&Instrument&$\lambda$& FWHM&Project \\
   & & $_{(\mic)}$& (arcsec)&ID\\
   \hline
   
   \emph{ISO}&CAM01&8.69&3&35700630\\  
   \emph{ISO}&CAM01&12.41&4.3&35700630\\  
   \emph{WISE}&W3&12&6&WISE Survey\\  
   \emph{ISO}&CAM01&13.53&4.7&35700630\\  
   \emph{WISE}&W4&22&12&WISE Survey\\  
   \emph{Spitzer}&MIPS&24&6&22508544\\ 
   \emph{IRAS}&HiRES&60&60&IRAS Survey\\   
   \emph{Herschel}&PACS&70&5.5&1342244-168/169\\  
   \emph{Spitzer}&MIPS&70&18&22508544\\ 
   \emph{Herschel}&PACS&100&6.7&1342196-767/768\\  
   \emph{Herschel}&PACS&160&11&1342196-769/770 \\ 
   \emph{Herschel}&SPIRE&250&18&1342187718\\ 
   \emph{Herschel}&SPIRE&350&25&1342187718\\ 
   \emph{Herschel}&SPIRE&500&37&1342187718\\ 

    \hline\\
\end{tabular}

\end{center}
\end{table}

To study the different emitting components in the nebula around
G79.29+0.46, we have retrieved  most of the spectroscopic and photometric data from the mid- to the far-IR present in the archives. The considered datasets are summarized in Tables \ref{tab:spectraproperties} and \ref{tab:mapsproperties}, which include data collected with  the: \emph{Herschel} Space Observatory,  \emph{Spitzer} Space Telescope, Infrared Space Observatory (ISO), Infrared Astronomical Satellite (IRAS), Wide-field Infrared Survey Explorer (WISE). In Table \ref{tab:mapsproperties} we also indicate the wavelength and the full width at half maximum (FWHM) for each dataset.

The \emph{Herschel} images come from the ``\emph{Herschel} imaging survey of OB Young Stellar objects'' (HOBYS, P.I. Motte F.) and from the ``Mass-loss of Evolved StarS" (MESS, P.I. Groenewegen, M.A.T.) programs, which used the instruments Photodetecting Array Camera and Spectrometer (PACS) and Spectral and Photometric Imaging Receiver (SPIRE). The PACS photometric data (from MESS) includes 100 $\mic$, while the SPIRE/PACS Parallel observation (HOBYS) provided the 70, 160, 250, 350, 500 $\mic$ data. The \emph{Herschel} images are presented here for first time, as well as the WISE \citep{2010Wright} data (which come from the Image Atlas), 
while the Multiband Imaging Photometer (MIPS) maps have already been published (\citealt{2010Kraemer,2010Esteban}; Paper 1). The ISO images were briefly discussed in the poster paper of \citet{1998Trams} and those IRAS (obtained with HiRes) by \citet{1994Higgs} and \citet{1996Waters}. 

From the \emph{ISO} data archive we also retrieved the spectroscopic datasets (see Table \ref{tab:spectraproperties}). The Long Wavelength Spectrometer (LWS) spectra were presented by \citet{1998Wendker}, but they provided only a qualitative discussion of
the data.  The Short Wavelength Spectrometer (SWS) spectra appeared for first time in the proceeding of
\citet{2008Morris}. The InfraRed Spectrograph (IRS) spectra were acquired with the Short-Low (SL), Short-High (SH) and Long-High (LH) modules. These data were already analysed by \citet{2010Esteban} but, despite the
interesting study proposed, these authors did not consider the
off-nebula dataset and hence they did not subtract the background, which instead
strongly affects the target continuum and line emissions. Moreover, they did not
apply corrections for extinction in their diagnostic study. Hence we propose a
new analysis. The positions of the IRS, SWS, LWS slits projected on the nebula around G79.29+0.46 are shown in the Appendix (see Fig. \ref{fig:slitirs} and \ref{fig:slit}).

\section{Analysis of the radio continuum emission}
\label{par:maps}
\begin{figure*}
\centering
 \includegraphics[width=\textwidth]{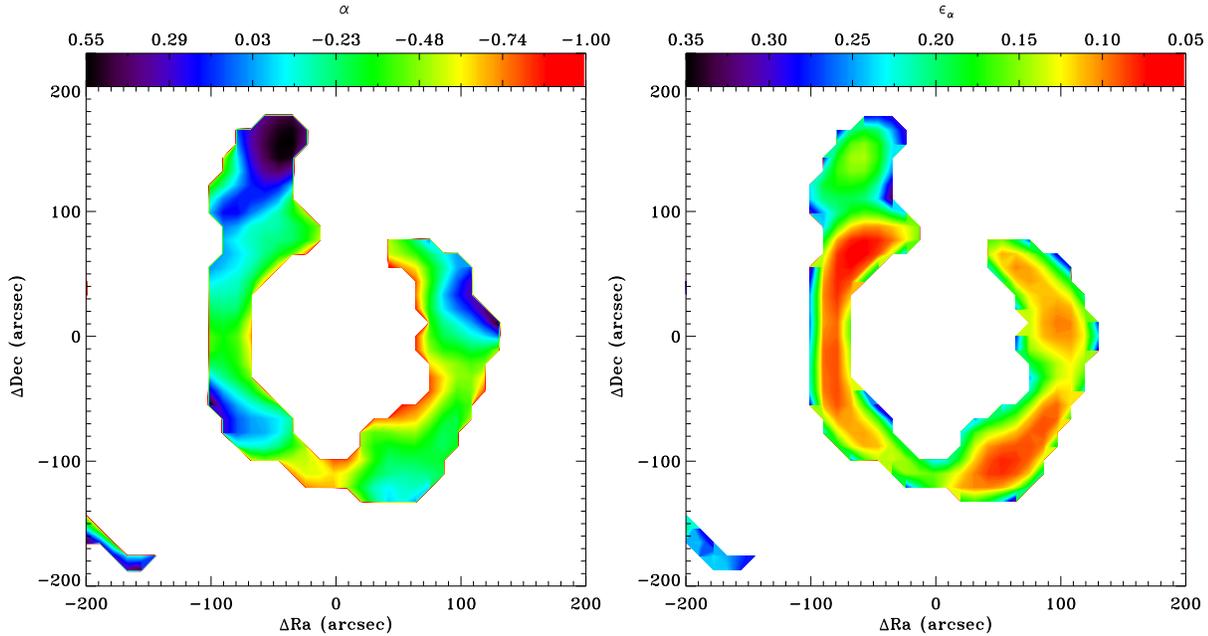}
\caption{Left: Spectral index map of G79 between 1.4 and 5 GHz. Right: Error
spectral index map.}
\label{fig:spix_g79}
\end{figure*}
\subsection{Morphology and analysis of the ionized gas}
\label{par:subsec}
The map at 1.4 GHz shows that the emission arises mostly from the
nebula (see left panel in Fig. \ref{fig:radiomaps}). The central object is not
detected. There are two possible causes: (1) the spectral index of stellar winds
($\alpha \sim0.6$) implies low flux density at low frequencies and (2) 
the confusion limit of the interferometer at this frequency. The ``spur-emission"
in the north-east of the nebula first noticed by \cite{1994Higgs} is visible. At 5 GHz it is possible to well distinguish the emission from a compact central object, as well as that from the nebula and from the spur (Fig. \ref{fig:radiomaps}, right panel). 
As already discussed in Paper 1, the highest resolution maps (at 5 GHz) reveal a highly structured texture of the nebula and the extragalactic source overlaid on
the nebula in the north-west part. The map seems filled with a diffuse emission, which has not any  counterpart with the outer IR shell, therefore we conclude that it is arises from the close HII region DR 15. On the contrary, at 8 GHz the array resolved out most of the extended emission from the
nebula, but the central object is still bright.

We have estimated the total flux density  over the radio shell from our final images, having the one at 1.4 GHz the best dynamical range and LAS, and that at 5 GHz being ``corrected'' for the flux-loss.  We have measured flux
densities by using CASA. At 5 GHz we also measured the background emission in different regions around the nebula and then we subtracted the average value from the total flux density. The flux density errors were determined as in equation 
\begin{equation}
\sigma=\sqrt{(RMS_{\mathrm{tot}})^{2}+(\sigma_{\mathrm{cal}}S_{\nu})^{2}} 
\label{equ:rms}
\end{equation} 
with: $RMS_{\mathrm{tot}}$ the rms-noise in the map integrated over the area covered by the source;  $\sigma_{\mathrm{cal}}$ the systematic relative error due to the flux calibrator (typically 3-5$\%$).  In the case of the 5 GHz measurements $RMS_{\mathrm{tot}}$ was replaced with the standard deviation in the considered regions multiplied by the area
of the source. The values found are: $S(1.4\,\rm GHz)=320\pm82\,\rm
mJy$ and $S(5\,\rm GHz)=323\pm126\,\rm mJy$. 
We highlight that the 5 GHz error is large because of the background emission, which is non-uniform and makes our estimate uncertain. We also note that the 1.4 GHz may be underestimated because not complemented with single-dish data.
The values found are lower than those obtained by \citet{1994Higgs} through interferometric and single-dish observations, but probably they did not subtract the background. This leads to the discrepancy with our measurements. We also compared our flux densities at both frequencies with those measured on the EVLA:D maps (with the best LAS) over the radio shell in order to verify their consistency. We found a good agreement.
\subsection{Spectral index}
\label{par:analys}

To better understand the nature of the radio nebula, we have created a spectral
index map between the EVLA:D$\&$C 1.4 and 5 GHz maps. We chose these maps because they
have comparable LAS (see Table \ref{tab:EVLAmaps}).
Before calculating the spectral index, we have re-gridded the map  at 5 GHz to the grid of 1.4 GHz. Then we have converted the maps (originally in unit of Jy
beam$^{-1}$) to Jy pixel$^{-1}$, dividing by the number of pixels contained in
 the area of a beam (at the HPBW). 
Then we deconvolved the maps from their initial beams
and reconvolved with a final beam $27\arcsec\times25\arcsec$. This operation is equal to
the convolution with a gaussian beam with
$\sigma^{2}=\frac{1}{8ln2}(HPBW^{2}_{i}-HPBW^{2}_{f})$. Finally, the spectral
index map was calculated discarding pixels below 5 $\sigma$, where the noise was evaluated
in the reconvolved images. 
Considering the error for the flux as in equation \ref{equ:rms} (in unit of $\rm Jy\,\rm pixel^{-1}$), with the uncertainty of the calibrator flux $\sigma_{cal}$ equal to 3\% at 5 GHz and 5\% at 1.4 GHz, we have also estimated the error
spectral index map. 
The spectral index and error maps are plotted in Fig. \ref{fig:spix_g79}.

Over the nebula, the average spectral index $<\alpha_{neb}>$ is
$-0.29\pm0.15$, which is consistent with the theoretical value ($-0.1$) for optically thin HII regions emitting free-free radiation. The lower limit of $<\alpha_{neb}>$ is -0.44, indicating
that the flux density at 5 GHz is underestimated because of flux-loss due to the interferometer. $\alpha_{neb}$ varies slightly in some interesting regions. For example, in the south-west part of the
nebula, close to the infrared dark cloud (IRDC) \citep{2010Kraemer}, $\alpha_{neb}=-0.18\pm0.14$, which is still consistent with optically thin
free-free but slightly higher than $<\alpha_{neb}>$. We can speculate that this is a
modified region due to shock (Paper 1). On
the ``spur-region'' the maximum value of $\alpha_{neb}$ is 0.55, with an error of
0.17. This value is usually consistent with stellar winds, but in this case it is
more likely due to emission from density clumps with a turn-over frequency
between the considered frequencies. This spur-region is not correlated with the IR emission of G79.29+0.46 and at the moment we are not able to establish if it belongs to the nebula or if it is really a foreground/background emission region.

To derive the nature of the central source we have determined its spectral index between 5 and 8 GHz from the corresponding flux densities. Therefore, we have measured its flux density by fitting a Gaussian around the
source in both the maps. As a result we found $S(4.96\,\rm GHz)=1.81\pm 0.08$ mJy and $S(8.46\,\rm GHz)=2.35\pm 0.19$ mJy, where the
error is determined as in equation \ref{equ:rms}. Therefore, we estimated the spectral index $\alpha_{centre}$ by using these measurements. The spectral index error was obtained adopting the error propagation for independent measures. 
We find $\alpha_{centre}=0.83\pm0.18$, which is consistent with a stellar wind
($\alpha=0.6$). The measurements we are comparing come from observations at different epochs (2005 and 2010 for the 8 and 5 GHz datasets, respectively) but, despite LBVs can suffer of changes in the mass-loss rate \citep[e.g. AG Car,][]{2001Stahl, 2002Vink, 2009Groh}, G79.29+0.46 has not yet shown any important variability, therefore we are confident that the two datasets are compatible. Previously, through their observations at 4.86 and 8.46 GHz of the central object,  \cite{1994Higgs} derived a spectral
index of $1.39 \pm 0.14$, which is steeper than the canonical value. This is very likely due to the fact that they obtained a larger beam at 8.46 GHz ($\sim 11\arcsec$) and therefore the flux measurement was contaminated by circumstellar or background emission. 
The authors pointed out that their spectral index is still consistent with mass-loss from a stellar object. With our estimate, now we can confirm that the central object
detected is a stellar wind from the candidate LBV G79.29+0.46. 
\subsection{Ionized mass}
\label{sec:em_mass}

Once known the nature of the radio nebula, we can estimate its average electron
density and its  ionized mass, since the optical depth $\tau_{\nu}$ for a nebula of H emitting free-free radiation and with a temperature about 10$^{4}$ K can be expressed as
\begin{equation}
 \tau_{\nu}=\int{\kappa_{\nu} dl}=8.24\times10^{-2}\left(\frac{T_{e}}{\rm
K}\right)^{-1.35}\left(\frac{\nu}{\rm 5\,GHz}\right)^{-2.1}\frac{EM}{\rm
pc\,cm^{-6}}
\end{equation}

where the emission measure (EM) is

\begin{equation}
 EM = \int_{0}^{s}{n_e^2 ds\,[\rm pc\,cm^{-6}]}
\label{equ:emmeasure}
\end{equation}
Therefore we can use the optical depth map (as determined from the radiative transfer equation in the case of  optically thin emission not absorbed along the
line of sight and considering the Rayleight-Jeans approximation for the Planckian, valid 
in the radio domain) to obtain the emission measure and hence the average electron
density from the radio measurements.  
For this analysis we refer to the final image at 5 GHz (Fig.
\ref{fig:radiomaps}), as it
covers all the angular scales of the considered region.

Before calculating the optical depth, we have subtracted the background emission (estimated by averaging the values found in different regions on the map) and we have converted the map from unit of
Jy beam$^{-1}$ to Jy pixel$^{-1}$, as before. 
We find an average $\tau_{\nu}$ of $\sim0.2$. Therefore, the average value for $<EM>$ over the shell is $680\pm270\,\rm pc\, cm^{-6}$. The error in  $<EM>$ takes into account the uncertainty of the flux density measurement. In order to
calculate also the electron density we need to establish the geometrical depth
of the nebula. We assume the geometry of a hollow sphere with average inner and outer radii of $<R_{in}>=80\arcsec$ (which corresponds to 0.66 pc at a distance of 1.7 kpc)
and $<R_{out}>=110\arcsec$ (0.9 pc) respectively. 
Therefore, we assume an average geometrical depth $<s>$ of 0.67 pc and we find an average electron density
 of $\sim 32\pm20\, \rm cm^{-3}$. Finally, the ionized mass is $1.51\pm0.96\,\rm M_{\odot}$ in the considered volume. 
Modelling the radio flux densities, \citet{1994Higgs} obtained a mass of 6.3 $\rm M_{\odot}$ at a distance of 1.7 kpc. These authors assumed a bigger volume for the shell (with inner and outer radii of $95\arcsec$ and $130\arcsec$ respectively)  and very likely did not subtract the background/foreground contribution from the nebula emission. Adopting their same values and the electron density ($\sim 45\pm36\, \rm cm^{-3}$) derived from the map not background-subtracted, we find a mass of $3.5 \pm 2.8\,\rm M_{\odot}$ which is lower than the value provided by \cite{1994Higgs}, but still consistent within the error.

\subsection{Current mass-loss and spectral type of the star}
\label{sec:sptype}
Since the emission from the central object is due to stellar wind (as shown in the previous
Section), from the measured flux density we estimate the mass-loss rate of the star,
following \citet{1975PF}:
\begin{equation}
 \dot{M}=6.7\times10^{-4}v_{\infty}S_{\nu}^{3/4}D^{3/2}(\nu\times
g_{ff})^{-1/2}\,\rm [M_{\odot}\,yr^{-1}]
\label{equ:massloss}
\end{equation}
where $S_{\nu}$=1.51 mJy is the observed radio flux density at 4.96 GHz,  $v_{\infty}=110\,\rm
km\,s^{-1}$ \citep{2000Voors} is the terminal velocity of the wind, $D=1.7\,\rm
kpc$ \citep{2010Esteban}. The free-free \emph{Gaunt} factor $g_\mathrm{ff}$  is  approximated as in 
\citet{1991Lei}: for a gas with electron temperature of $T_e=10000\,\rm K$ at 4.96 GHz, $g_\mathrm{ff}=5.08$. 
Therefore, we find that the star is loosing mass with a rate of $(1.4\pm0.2)\times10^{-6}(\frac{v_{\infty}}{110\,\rm
km\,s^{-1}})\ (\ \frac{D}{1.7\,\rm kpc})\ ^{3/2} f^{1/2}\,\rm M_{\odot}\,yr^{-1}$ (where $f$ is the filling factor). 

Once we have evaluated the ionized mass in the nebula (as in the previous
subsection), we can provide an estimate of the ionizing photons flux from the
star F$_{UV}$, as

\begin{equation}
 F_{UV}=\frac{M_{ion}<n_e>\beta_2}{m_{p_{[\rm M_{\odot}]}}}
\label{equ:fuv}
\end{equation}
approximation valid for an ionization bounded nebula (as found in Paper 1). Considering for the recombination coefficient $\beta_2$ and the proton mass $m_{p_{[\rm
M_{\odot}]}}$ the standard values, we find $\log
F_{UV}=46.24$, which corresponds to a spectral type from B2 to B1 for a supergiant I
\citep{1973Panagia}. \citet{2000Voors} proposed a mid-B supergiant after analysis of optical and near-IR spectra of the central object. However, the value we found must not be considered as the current spectral type of the star, as the recombination time of the gas is much longer (tens thousands years) than the typical LBV variability cycles (decades). This implies that the radio emission from the shell is not sensitive to changes in the stellar parameters. The spectral type derived with this method must be considered an average value over $\sim10^{4}$ years.

\section{Analysis of the IR continuum emission}
\label{par:infrared}
\begin{figure*}
 \centering
\includegraphics[width=.7\linewidth]{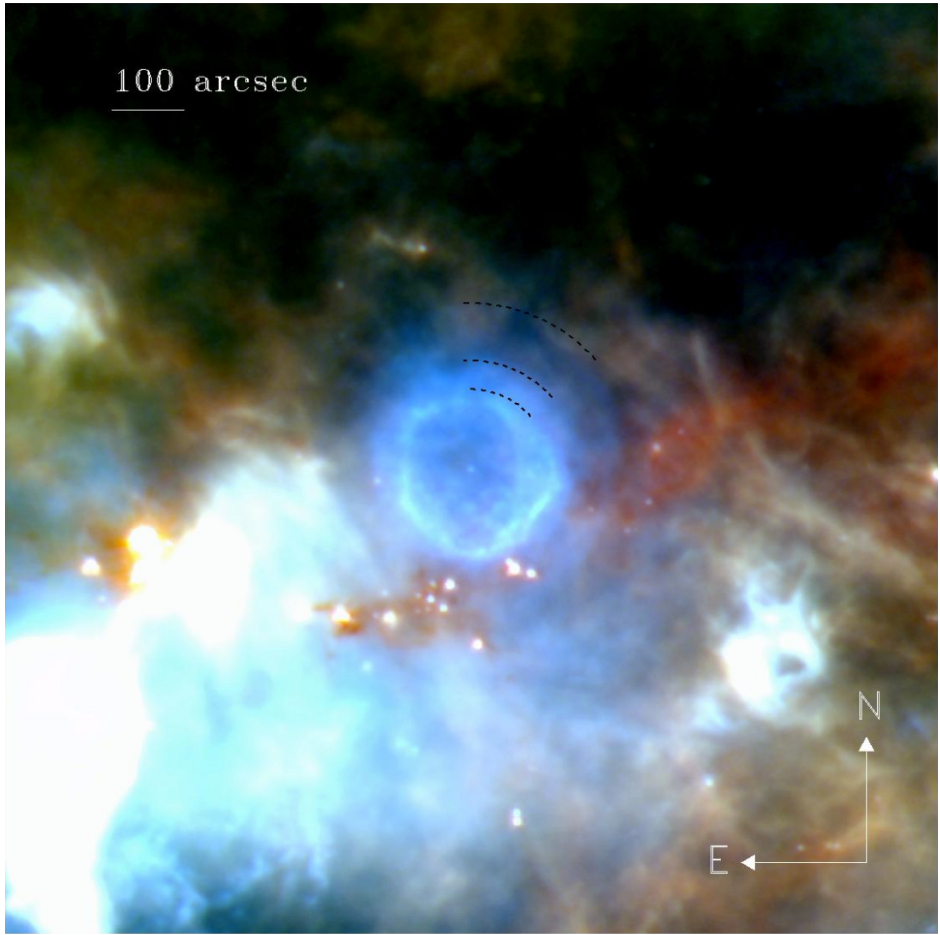}
\caption{Three-colour image of the nebula around G79.29+0.46 from the \emph{Herschel} data at 70 $\rm \mu m$ (blue), 100 $\rm \mu m$ (green) and 160 $\rm \mu m$ (red). Edges of the three shells are highlighted with arcs.}
\label{fig:rgb}
\end{figure*}

\begin{table}
\centering

  \caption{Total flux densities of the IR nebula}
\label{tab:totfluxes}   

  \begin{tabular}{lcl}

\hline
   $\lambda$&Total &  Instrument \\
   &Flux density$^{*}$&\\
   $_{(\mic)}$ & $_{(\rm Jy)}$&\\

   \hline
   8.69&7.74$\pm$1.55$^{a}$&ISOCAM01\\
   11.48&11.94$\pm$2.39$^{a}$&ISOCAM01\\
   12&14.8$\pm$2.0$^{a}$ &WISE-W3\\
   12.41&8.9$\pm$1.8$^{a}$&ISOCAM01\\
   13.53&9.11$\pm$1.82$^{a}$&ISOCAM01\\
   15.96&20.6$\pm$4.1$^{b}$&ISOCAM01\\
   22&166.1$\pm$17.0$^{c}$&WISE-W4\\
   24& 165.8$\pm$17.0$^{c}$&MIPS\\
   25&194.9$\pm$20.0$^{c}$&IRAS-HiRES\\
   60&725.6$\pm$218.0$^{c}$&IRAS-HiRES\\
   70&552.2$\pm$110.0$^{c}$&MIPS\\
   70&583.7$\pm$117.0$^{c}$&PACS\\
   100&500$\pm$350$^{c}$&PACS\\
   160&31.8$\pm$10.0$^{d}$&PACS\\
   \hline
\end{tabular}
\footnotesize

    $^{*}$ Errors include only calibration uncertainties\\ 
    $a$ Inside a circle with radius 1.96 arcmin\\
    $b$ Inside a circle with radius 2.2 arcmin\\
    $c$ Inside a circle with radius 3.6 arcmin\\
    $d$ Inside a circle with radius 3 arcmin\\
\end{table}

\subsection{Morphology of the infrared nebula from the \emph{Herschel} images}

\begin{figure*}
\centering
 \begin{minipage}{1\linewidth}
 \includegraphics[width=\textwidth]{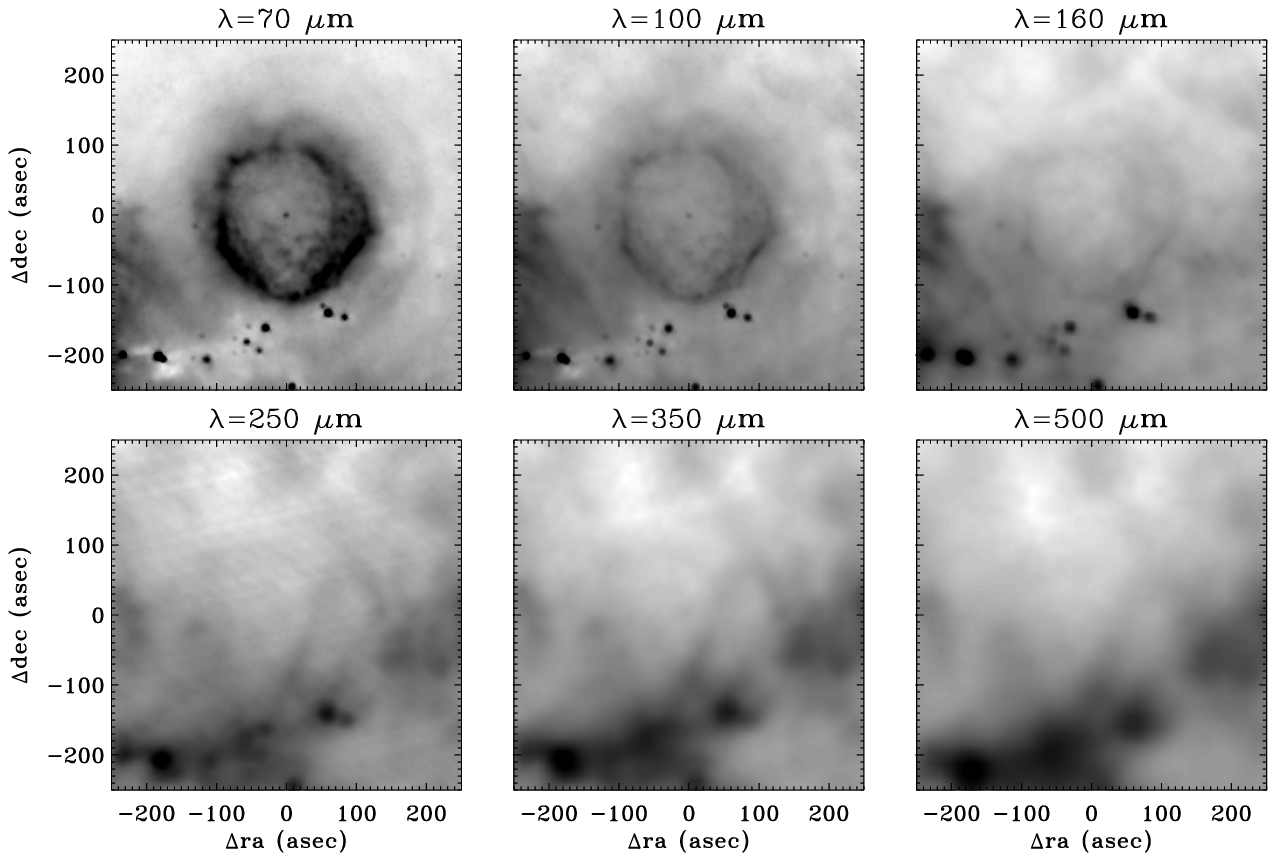}
 \caption{The \emph{Herschel} images of G79.29+0.46. Panel (a): PACS $70\,\mic$ (left), $100\,\mic$  (centre) and $160\,\mic$ (right).
 Panel (b): SPIRE $250\,\mic$ (left), $350\,\mic$  (centre) and $500\,\mic$ (right). 
 For  each map, the field of view is about $4^{\prime} \times 4^{\prime}$ centred on the central source  position.}
\label{fig_herschel}
 \end{minipage}
\end{figure*}

The high-resolution \emph{Spitzer} images have been already described in different works, where the morphology of the infrared nebula has been pointed out: it consists in multiple shells, indicative of mass-loss episodes at different epochs. These previous works showed that there are at least two populations of dust: the hot one
emitting in the InfraRed Array Camera (IRAC) bands, and the warm one emitting in the
MIPS bands. The hot dust ($\sim1500$ K) is distributed in the inner ring (with peak at
about $\pm100\arcsec$), mostly cospatial with the ionized gas (\citealt{2010Kraemer}; Paper 1). The warm dust ($\sim90-350$ K) emitting at longer wavelengths ($>24\, \mic$) is more extended in the inner ring than the hot dust \citep{2010Kraemer} and presents a further component in the outer shell  \citep[peaked at about $\pm200\arcsec$,][]{2010Kraemer,2010Esteban, 2011UmanaG79}. A central object, on the position of the star, is  detected up to 24 $\mic$.

Fig. \ref{fig:rgb} is a three-colour image obtained superimposing the \emph{Herschel}/PACS images at $70\,\mic$ (in blue), $100\,\mic$ (green) and $160\,\mic$ (red). We set up the absolute flux scale of the \emph{Herschel} images by using the standard calibration method
developed by \citet{2010Bernard}. The $70\,\mic$ and $100\,\mic$ images have a resolution comparable to that of the \emph{Spitzer}/MIPS image at $24\,\mic$,
while at $160\,\mic$ it is comparable to that of the MIPS image at $70\,\mic$ (see Table \ref{tab:mapsproperties} and Fig. \ref{fig_herschel}) . The $70\,\mic$ and $100\,\mic$ maps show the overall morphology of the
$24\,\mic$ emission, probing the warm dust distributed in the inner and outer shells. However, we note that the broad inner shell reveals actually another component, peaked at about $\pm150\arcsec$, leading to a total of three dusty shells (as indicated with arcs in Fig. \ref{fig:rgb}). The first two shells (the wide bright one peaked at $\pm100\arcsec$ and the close faint peaked at $\pm150\arcsec$) have a counterpart also at $160\,\mic$. At this wavelength there are also hints of the outer shell, but the background is high and confusion becomes dominant. For the photometric measurements, in this case we will
concentrate only on the two inner shells (lower limit). We also note that the central object is quite bright at $70$ and  $100\,\mic$ and weaker, but still visible, at $160\,\mic$.

In the SPIRE bands ($250\,\mic$, $350\,\mic$ and $500\,\mic$) the central object is not detected, the nebula
is less bright, and the IRDC in the south appears in emission (as evident in the bottom panels of Fig. \ref{fig_herschel}).
At these wavelengths there are hints of arcs in the
position of the inner shell, but the IRDC becomes brighter and contaminate the
shell emission. For this reason, in our analysis we will refer only to the
$70\,\mic$, $100\,\mic$ and $160\,\mic$ images. The absence of further nebular features in the position of the nebula may exclude the presence of a cold dust component, even if it could be not detected because too faint relative to the surrounding interstellar medium.

\subsection{Analysis of the infrared emission}

\label{sec:irimages}

\begin{table}
\centering

  \caption{Total flux densities of the IR central object}
\label{tab:totfluxescentral}   

  \begin{tabular}{lcll}

\hline
   $\lambda$&Total &  Instrument & Comment\\
   &Flux density$^{*}$&&\\
   $_{(\mic)}$ & $_{(\rm Jy)}$&&\\
   \hline

   1.235&2.73$\pm$0.02&2MASS&2MASS Catalogue\\
   1.662&7.81$\pm$0.04&2MASS&2MASS Catalogue\\
   2.159&12.37$\pm$0.04&2MASS&2MASS Catalogue\\
   3.4&10.17$\pm$4.51&W1&WISE Catalogue\\
   3.6&10.57$\pm$1.05&IRAC&Cyg-X Legacy Catalogue\\
   4.5&9.40$\pm$1.00&IRAC& Cyg-X Legacy Catalogue\\
   5.8&6.82$\pm$0.50&IRAC&Cyg-X Legacy Catalogue\\
   8.0&4.60$\pm$0.40&IRAC&Cyg-X Legacy Catalogue\\
   12.0&2.32$\pm$0.05&W3&WISE Catalogue\\
   22.00&0.93$\pm$0.04&W4&WISE Catalogue\\
   23.99&1.08$\pm$0.02&MIPS&Cyg-X Legacy Catalogue\\
   24.0&1.01$\pm$0.05&MIPS&This work\\
   70.0&0.44$\pm$0.04$^{*}$&PACS&This work\\
   100.0&0.30$\pm$0.03$^{*}$&PACS&This work\\
   160.0&0.11$\pm$0.03$^{*}$&PACS&This work\\

   \hline
\end{tabular}
\footnotesize

    $^{*}$ Errors include only calibration uncertainties\\ 
\end{table}

We have measured flux densities of the nebula around G79.29+0.46 by using the package \emph{Skyview}\footnote{http://www.ipac.caltech.edu/skyview/
}, release 3.6. We remind
the reader that all the images we used are summarized in Table \ref{tab:mapsproperties}. Considering the circular shape of the nebula, flux densities were evaluated in
circles enclosing the emitting regions to the specific wavelength (as indicated in Table  \ref{tab:totfluxes}). 

To take into account the strong diffuse emission around the nebula, we evaluated the background in annuli close to the nebula and we subtracted it from
the integrated flux densities. As already mentioned, the background is important and extremely variable, especially at the longer wavelengths ($\geqslant 24\, \rm \mic$), and this causes a high uncertainty in its determination ($\sim$15--25 $\%$). This uncertainty, integrated over the emission region, causes errors up to $\sim50$--$70\%$ on the flux densities. In Table  \ref{tab:totfluxes} we report flux densities and their error at all the wavelengths. Note that in the Table the flux densities error includes only calibration uncertainties.

\subsection{The nature of the IR central object}
\begin{figure*}
\centering
\includegraphics[width=.5\textwidth, angle=90]{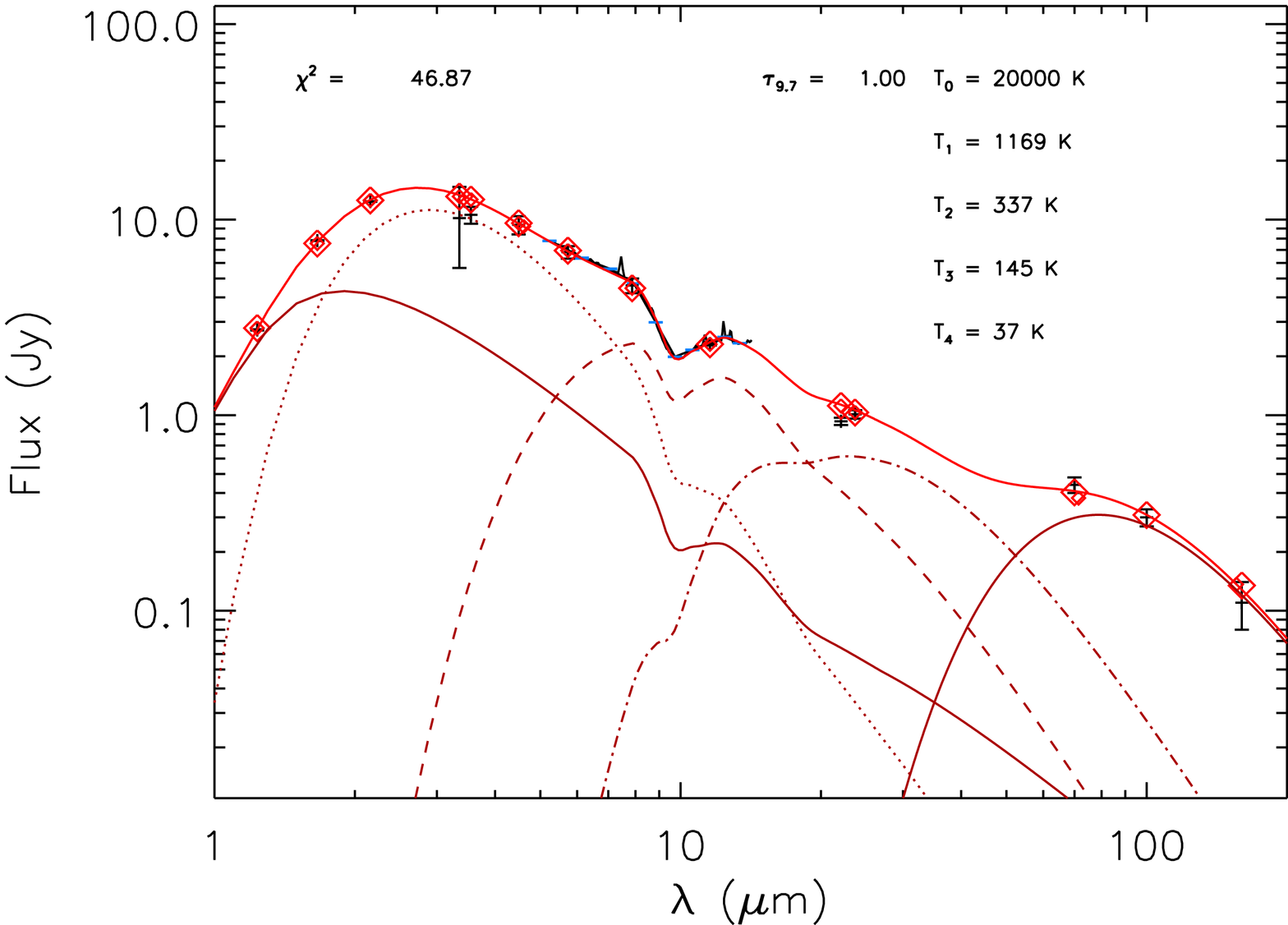}
\caption{Five-components fitting of the central source SED. Red diamonds are the best fit values and show both the star (black-body) and four dust components ($\nu^2 B_{\nu}(T)$ grey-bodies). Black crosses are the observed values. The IRS/SL spectrum is shown in blue. Derived parameters are shown as text (temperature ``$T_i$" for each component, reduced $\chi^2$,  and optical depth at 9.7 $\rm \mu m$).}
\label{fig:centralsource}
\end{figure*}

Hot stars like G79.29+0.46 are not expected to be detected at the longer wavelengths. However, at the position of the star, the \emph{Herschel} images show a compact emission at 70, 100 and 160 $\mic$. To understand the nature of this emission, we consulted the IRSA point source catalogues and derived the near-IR and mid-IR flux densities of the star from the 2MASS J, H, K and the WISE magnitudes\footnote{The 2MASS \& WISE magnitudes were transformed into flux densities using their well known "zero values" (see their release notes).}. We also included the IRAC and MIPS photometric measurements from the \emph{Spitzer} Legacy Survey of the Cygnux-X complex \citep{2007Hora}. For flux densities at the longer wavelengths (PACS bands) we used our own measurements, obtained by using an aperture of 6 arcsec and the following aperture correction: 1/(0.701,0.672,0.438) at 70, 100, 160 $\mic$ respectively. The mentioned flux densities for the central source are listed in Table \ref{tab:totfluxescentral}.

We have hence fitted  the infrared photometric data with black-body and grey-body  functions, following the method used in \citet{2011Flageybis} to constrain the nature of other hot stars surrounded by mid-IR shells by means of IRS spectra and near- to mid-IR photometric measurements. In the fitting we also included the IRS/SL spectrum (Section \ref{sec:irspectra}) taken on the position of the star. We note that the photometric data match very well the IRS data without any scaling factor, despite the background is evaluated in different regions. 
We find that a minimum of five components are required to obtain a decent fit of the complete spectral energy distribution (SED). We impose  in particular a black-body for the central star and $\nu^{2}B_{\nu}$ grey-bodies for the other components. We set the effective temperature of the star $T_{eff}$ at 20000 K. The free-parameters are: the four dust temperatures from $T_{1}$ to $T_{4}$, the amplitudes of the black/grey-bodies, and the extinction expressed as a depth $\tau_{9.7}$.  Each of these parameters can vary within a defined range: $T_{eff}$ from 20000 and 24000 K; $T_{1}$ from 500 to 2500 K; $T_{2}$ from 150 to 500 K; $T_{3}$ from 50 to 150 K; $T_{4}$ from 30 to 100 K; $\tau_{9.7}$ from 0 to 10.
A best-fit is shown in Fig. \ref{fig:centralsource}. 

The derived parameters are: the temperatures ($\simeq$ 1170, 340, 145, 40 K) of the four dust components; the depth at 9.7 $\mic$ (1); the reduced $\chi^2$ (46.87). However, we draw the reader's attention to the fact that, because of the many degrees of freedom and possibly degeneracy between them, we find many similar best fits with slightly different final parameters, depending on the initial guess of the fit. The four dust component temperatures should thus be seen as rough estimates. We suggest that there is a range of temperatures, well described by the range explored by $T_{1}$ to $T_{4}$, in a circumstellar envelope close to the central star. The temperature of the star is not well constrained, as shown in the Figure, as the black-body contributes very little to the J band only. 

 The IRS spectrum\footnote{For this analysis we do not apply extinction correction.} (described in Section \ref{sec:irspectra}) gives a strong evidence that an absorption feature at 10 $\mic$ due to astronomical silicates is present. Therefore, we use the depth $\tau_{9.7}$ as obtained in the fit to determine the visual extinction $A_{V}$ toward the star, by means of the extinction curve determined by \citet{1990Mathis}. We find $\tau_{9.7}=1$ (with an uncertainty of about 10$\%$), therefore we obtain $A_{V}$=18.6 in the case of a constant total-to-selective parameter $R_V=3.1$ (typical value for small grains 
in the Milky Way). 

Visual extinction of G79.29+0.46 has been determined by various authors and optical/near-IR observations of the source have provided diverse values for $A_{V}$. For example,  \citet{1994Higgs} found a value of 16$\pm$1 mag by comparing the shape of the observed red continuum with reddened model continua of early type stars. \citet{1999Trams} derived $A_{V}=11.9$ mag by studying the infrared properties of G79.29+0.46. \citet{2000Voors} derived a visual extinction of 14.9 mag by means of JHK spectra, assuming an effective temperature of 18000 K and the extinction law by \citet{1991Torres}. \citet{2000Voors} also measured the extinction from the Diffuse Interstellar Bands (DIBs), by using a relation between the strengths of these bands and $E(B-V)$ valid in several OB associations in Cygnus. In this case they found  $A_{V}\sim6.4$ mag which is much less than the value derived previously. They explained the discrepancy between the two methods as due to the fact that the DIBs do not give measure of the extinction 
due to molecular clouds, 
while G79.29+0.46 is very close to the obscured star-forming region DR 15. Because of the many degrees of freedom of our model, we can not assess which method among these is the best for determining the extinction towards the star. For the moment, we assume that $A_{V}$ must range between $\sim$ 6 and $\sim$ 18.6 mag (see Section \ref{sec:extinction} for more details about extinction).

\subsection{IR physical properties of the extended nebula}

Following a simple method described in \citet{1998Dayal} and in
\citet{2010Umana}, we also estimated the dust temperature and mass for the dust components in the extended nebula. The average dust temperature can be derived interpolating a Planckian distribution between two
observed brightnesses. Assuming that the infrared continuum emission observed is due to thermal dust\footnote{Our spectroscopic analysis in next Section will show that there is no evidence of the presence of Polycyclic Aromatic Hydrocarbons (PAHs). However, emission lines can contaminate the continuum emission at wavelengths $<15\,\rm \mic$.} and is not absorbed by colder dust along the line of sight, from the radiative transfer equation and the optical depth, we can derive the column density, $\rho l$ and, therefore, the mass of the nebula, as 
\begin{equation}
 M=\rho l\Omega D^2
\end{equation}
where $\Omega D^2$ is the physical area covered by the emitting region. For the analysis of the hot dust we consider the ISOCAM maps at 8.69 
$\mic$ and at 13.53 $\mic$, which have the same resolution. 
We do not consider in this analysis the IRAC maps because too much contaminated by the saturated star emission. 
From the measured ISOCAM flux densities in Table \ref{tab:totfluxes} we subtracted the
stellar component, through aperture photometry. After this subtraction, we estimated for the nebula the following flux
densities: $S(8.69 \mic)= 6.5\pm1.3\,\rm Jy$ and $S(13.53 \mic)= 8.3\pm1.7\,\rm Jy$. For the analysis of the warm dust component we refer to the
\emph{Spitzer}/MIPS 24 $\mic$ and the \emph{Herschel}/PACS 70 $\mic$ maps, given
their similarity in resolution. The nebula flux densities in this case are $S(24 \mic)= 165\pm17\,\rm Jy$  and $S(70 \mic)= 583\pm117\,\rm Jy$ after subtraction of the central source contribution. 

\begin{table*}
\small
\begin{center}
\caption{Assumed chemical composition, grain size and absorption coefficient;
derived optical depth, temperature and mass of the dust, where $\lambda_1$ and  $\lambda_2$ are 8.69 $\rm \mu m$ and 13.5 $\rm \mu m$ respectively (in the case of hot dust) and 24 $\rm \mu m$ and 70 $\rm \mu m$ (warm dust).}
\label{tab:tempmass}

\begin{tabular}{lccccccc}
   \hline
  
Composition&$\kappa_{\lambda_1}$&$\kappa_{\lambda_2}$&$\tau_{\lambda_1}$&$\tau_{\lambda_2}
$&T&$M_{\lambda_1}$&$M_{\lambda_2}$\\
   and size ($\mic$)&${_{(\rm cm^2\,g^{-1})}}$& ${_{(\rm
cm^2\,g^{-1})}}$&&&${_{(\rm K)}}$&${_{(\rm M_{\odot})}}$&${_{(\rm M_{\odot})}}$\\
   \hline
                                   \multicolumn{8}{c}{\emph{Hot dust}}\\

                       \hline
  Silic. ($0.01,0.1$)&762&743&$(2.5\pm0.5)\times10^{-9}$&$(3.5\pm0.7)\times10^{-9}$&$518\pm210$&$(4.6\pm0.9)\times10^{-8}$&$(6.5\pm1.3)\times10^{-8}$\\
Silic. ($1$)&807&920&$(1.7\pm0.3)\times10^{-9}$&$(2.6\pm0.5)\times10^{-9}$&$587\pm238$&$(2.9\pm0.6)\times10^{-8}$&$(4.0\pm0.8)\times10^{-8}$\\
   Graph. ($0.01$)&390&230&$(1.1\pm0.2)\times10^{-8}$&$(9.5\pm2.0)\times10^{-9}$&$357\pm145$&$(3.9\pm0.8)\times10^{-7}$&$(5.8\pm1.2)\times10^{-7}$\\
   Graph. ($0.1$)&476&257&$(1.4\pm0.3)\times10^{-8}$&$(1.1\pm0.2)\times10^{-8}$&$336\pm136$&$(4.2\pm0.8)\times10^{-7}$&$(6.3\pm1.3)\times10^{-7}$\\

   Graph. ($1$)&1980&985&$(1.9\pm0.4)\times10^{-8}$&$(1.4\pm0.3)\times10^{-8}$&$319\pm129$&$(1.30\pm0.3)\times10^{-7}$&$(2.0\pm0.4)\times10^{-7}$\\
\hline
                                   \multicolumn{8}{c}{\emph{Warm dust}}\\

                       \hline
  Silic. ($0.01, 0.1,
1$)&647.7&69&$(4.0\pm0.4)\times10^{-4}$&$(4.2\pm0.8)\times10^{-5}$&$59\pm18$&$0.032\pm0.006$&$0.032\pm0.006$\\
   Graph. ($0.01$)&300&107& $(6.1\pm0.6)\times10^{-5}$ &$(2.2\pm0.4)\times10^{-5}$&$72\pm22$&$0.011\pm0.002$&$0.011\pm0.002$\\
 Graph. ($0.1$)&309&109&$(6.2\pm0.6)\times10^{-5}$&$(2.2\pm0.4)\times10^{-6}$&$72\pm22$&$0.010\pm0.002$&$0.010\pm0.002$\\
   Graph.
($1$)&600&300&$(3.6\pm0.4)\times10^{-5}$&$(1.8\pm0.4)\times10^{-5}$&$77\pm23$&$0.0031\pm0.006$&$0.0031\pm0.006$\\
   \hline

\end{tabular}
\end{center}

\end{table*}
In the case of the hot grains, with an average temperature of $\sim 552\, \rm K$ (silicate) or $\sim 337\, \rm K$ (graphite),
the resulting average mass is very small ($\sim10^{-7}\,\rm M_{\odot}$) and hence
negligible. The warm dust has an average temperature of $59\, \rm K$ (graphite) or $74\, \rm K$ (silicate) and
results to be more massive ($3.2\times10^{-2}\,\rm M_{\odot}$ or $8.0\times10^{-3}\,\rm M_{\odot}$, for graphite or silicate, respectively). In all the cases
the dust is optically thin. The temperature and mass  are summarized in Table
\ref{tab:tempmass}, as derived for both the hot and warm dust components. Note that they depend on the chemical composition and are obtained assuming single grain sizes as indicated. Uncertainties take into account only the flux densities errors. By using a dust model to fit the IRAS images, \citet{1996Waters} provided a lower limit for the inner
shell temperature of about 65 K. They also derived a dust mass of $0.14 \,\rm M_{\odot}$. To understand the discrepancy with our dust mass (few $10^{-2} \,\rm M_{\odot}$), we have used the model described by \citet{1996Waters} and \citet{1997Izumiura}, based basically in the integration of the mass-loss rate. By using our parameters, we find a good agreement between the two methods. Therefore, the discrepancy is mostly due to the assumed parameters, in particular to the mass-loss rate and the outer radius of the shell, very likely  overestimated in \citet{1996Waters} because of the IRAS images bad resolution. Modelling the $8\,\mic$ and $24\,\mic$ emission from the \emph{Spitzer}/MIPS
image with a radiative transfer code, \citet{2010Kraemer} found slightly higher temperatures (87--108 $\rm
K$), in the hypothesis of a carbon-rich nebula with amorphous carbon grains, but they considered a larger distance for G79.29+0.46 in their calculations ($D=3\,\rm kpc$).

Mixed composition of silicate and graphite in dust has been observed in LBVNe \citep[e.g. Wray 17-96,][]{2002Egan}. Similarly, we assume that the nebula around G79.29+0.46 has a mixed chemistry and, considering the average mass for silicate and that for graphite, we derive an average total dust mass of $0.02 \,\rm M_{\odot}$. This leads to a gas-to-dust ratio of $\sim80$, being the ionized mass $1.51 \,\rm M_{\odot}$ (as estimated in Section \ref{sec:em_mass}). The value found is lower than the galactic interstellar case (100), but still consistent. However, this must be considered a lower limit, as we are not taking into account the neutral mass very likely distributed in the outer shell.

\section{Analysis of the mid-IR spectra}
\label{sec:irspectra}
In this section we present our analysis of the archival mid- and far- IR spectra collected by the
spectrometers LWS and SWS on board of \emph{ISO} and by the instrument IRS on
board of \emph{Spitzer}. The position of the spectrometers during the \emph{ISO} and \emph{Spitzer} observations are shown in the Appendix (see Fig. \ref{fig:slitirs} and \ref{fig:slit}). In the case of IRS spectra, we refer to the data obtained after the background
subtraction. Data taken at the same channel were also merged in order to work
with a spectrum per module. The mentioned spectra are 
shown in the Appendix (see Fig. \ref{fig:irs-spectra}, \ref{fig:sws-spectra} and  \ref{fig:lws-spectra}). 
Since G79.29+0.46 is a heavily reddened object, before to analyse the observed
lines, the fluxes must be corrected for extinction along the line of sight. 
For the analysis, we will refer only to
the corrected fluxes.

\subsection{Correction for extinction}
\label{sec:extinction}
Recent studies on the Cygnus-X  region report a median $A_{V}$ of 5.6 \citep{2003Hanson} and 5.7 \citep{2007Albacete}, respectively based on near-IR and X-ray observations.  To correct for interstellar extinction the IR spectra of the nebula, we assumed that the visual extinction for the nebula must be similar to that determined from other stars in the same region. These values are much lower than that obtained by fitting the infrared emission of the central source (Section \ref{sec:irimages}). However, since we can not estimate the intrinsic extinction of the source due to the shell and the dusty central object, we consider for the correction the lowest value 5.6, keeping in mind that the derived line fluxes may be lower limits. However, in the analysis, we will discuss the effects of taking $A_{V}=18.6$ (Section \ref{sec:diagnostic}). Therefore, we used $A_{V}=5.6$ and the extinction law determined by \citet{1990Mathis} for a constant total-to-selective parameter $R_{V}=3.1$. 

We measured line fluxes before and after the extinction correction. To derive the fluxes, we used the ISAP package version 2.1 and  fitted the lines
by means of single Gaussians, after having estimated the baseline, usually of the
first-order but in some cases of the second or third-order. We report intensities with a signal-to-noise ratio (S/N) usually larger than 5. The
errors are due to uncertainties in the background fitting procedure and do not
take into account the absolute calibration errors (better than 5$\%$ for IRS and SWS, and than 7$\%$ for LWS). An additional uncertainty component, not included because unknown, arises from the extinction curve. 
Line fluxes are listed in the Appendix (Tables \ref{tab:swslines}, \ref{tab:lwslines} \ref{tab:ch0lines}, \ref{tab:ch1lines} and \ref{tab:ch3lines}).

\subsection{Detected lines and shock-versus-PDR diagnostics}

The mid- and far-IR spectra of G79.29+0.46 are rich in ionic and atomic fine-structure
lines (e.g., [CII], [OI], [SiII], [FeII]), similarly to other LBVs \citep[e.g.
HR Car,][]{2009Umana}. In G79.29+0.46 this was first pointed out by
\citet{1998Wendker} and \citet{2010Esteban}. However, after the background
subtraction from the IRS spectra, we find some differences with \citet{2010Esteban}. For instance, some emission-lines disappear. In particular the [FI] 24.75 $\mic$ and [SiII] 34.80  $\mic$ lines. [ArII] (or [NeII]) 6.97 $\mic$, [ArIII] 8.99 $\mic$, [ArIII] 21.82 $\mic$, 
 [NeIII] 10.88 $\mic$ and [NiII] 10.68 $\mic$ are also not present\footnote{However, these lines detected by \citet{2010Esteban} may be present in other IRS datasets, which we have not re-analysed because off-nebula observations were not available.}. [NiII] 10.68 $\mic$ is visible, but it does not reach $3\sigma$. We also do not detect the H$_{2}$ transitions at 17.03 and 28.21 $\mic$. All these lines surely arise from the interstellar medium.

\citet{2010Esteban} attributed emission-lines in the band of the Unidentified
Infrared (UIR) features (centred at $3.3, 6.2, 7.7, 8.6, 11.3\,\mic$) to
Polycyclic Aromatic Hydrocarbons (PAHs). After subtracting the background
emission, we find that these features in the IRS/SL disappear. Therefore the emission in the UIR bands observed by \citet{2010Esteban} is
interstellar. In the Appendix, Fig. \ref{fig:slitirs} gives an indication that the mentioned lines belongs to the interstellar medium. Conversely, the lines showed in Fig. \ref{fig:irs-spectra} arises from the nebula (as shown in Fig. \ref{fig:slitirs}).
According to \citet{2010Esteban}, we confirm the absence of [SI]
25.2 $\mic$. 

\emph{ISO}/LWS spectra have strong [OI] 63 $\mic$, [OIII] 52 and 88 $\mic$, [NII] 122 $\mic$ and [CII] 157  $\mic$ line emissions (we remind the reader that unfortunately ISO off-nebula observations are not available in the archives).
[NII] 122 $\mic$ line is indicative of high-ionization regions. Together with
the [OIII] 88  $\mic$, it is brighter inside the inner shell, along the
east-west direction. The presence of both transitions in the
outer shell also indicates an extended nature of the ionized gas. We can not exclude
that it is interstellar diffuse gas, because off-nebula observations are not
available. However, the line fluxes vary across the nebula, hence at least some of it may belong to the source.

 Looking at the [OI] 63.2 $\mic$ line fluxes detected with \emph{ISO}/LWS, we
can notice that they are very similar in flux over the nebula. The [OI] emission
is typical of shocked regions or of photo-dissociation regions (PDRs). This line is the main cooling
transition in the dense post-shocked region of dissociative shocks
\citep{2010Vanloon,2001Giannini}. When the [OI] line is fainter than the [OIII]
line, it indicates absence of strong interactions \citep{2010Vanloon}. In this case 
[OI] line forms in PDRs, while [OIII] is typically strong in high-ionized
diffuse gas \citep{2002Mizutani}. The comparison in the LWS spectra
between these two transitions seems to exclude the presence of shocked regions, however background observations are needed in order to be sure that all the line emission arises from the nebula.
Another indicator of shocked regions or PDRs is the line pair [OI]
63$\mic$/[CII] 157.7 $\mic$ \citep{1985Tielens}. Strong [CII] emission is 
typical of PDRs. In particular, when the ratio is less then 10, then the
presence of a PDR is favoured. From analysis of this ratio in our LWS it seems excluded the presence of shocked regions in the area covered by the
observations (Fig. \ref{fig:slit}). It is clearly evident that a PDR is present between the HII
region, probed by free-free emission, and the
molecular cloud, probed by rotational CO line emission \citep{2008Rizzo}.
However, since for these spectra we do not have background datasets, we can not
exclude that this gas belongs to the diffuse interstellar medium. For this reason, we neglect the PDR mass in the total nebula mass estimation.

[OIII] 51.8 $\mic$ is particularly bright in the north-east part of the second
shell, next to the radio ``spur-emission" (see Fig. \ref{fig:radiomaps}). Analysis of the radio spectral index in this region showed us a value
(0.55) different from optically thin free-free emission. Excluding excitation by
a shock, these higher [OIII] emission in the nebula outskirts may be attributed
to density clumps.

\subsection{Electron density from mid-IR line ratios}
\label{sec:diagnostic}

\begin{table*}
\begin{center}
\caption{Line ratios and electron densities.}
\label{tab:OIIIlineratios}

\begin{tabular}{lccccc}
   \hline
   Slit&Size&Relative&Line pair& Line ratio&n$_e$ \\
   &(arcsec)$^2$&position$^{a}$ (arcsec)$^2$& &&$_{(\rm cm^{-3})}$\\
   \hline
   IRS-SH/LH&4.7$\times$11.3&0&18.5$\mic$/33.4$\mic$&1.45&$\sim$1000\\
   SWS-901&33$\times$20&0&18.5$\mic$/33.4$\mic$&0.74&$\sim$250\\
   LWS-003&84$\times$84&0&51.8$\mic$/88.4$\mic$&0.73&150-160\\
   LWS-008&84$\times$84&100 W&51.8$\mic$/88.4$\mic$&0.55&$<$100\\
   LWS-014&84$\times$84&100 E&51.8$\mic$/88.4$\mic$&0.53&$<$100\\
   LWS-011&84$\times$84&200 W&51.8$\mic$/88.4$\mic$&0.67&$\sim$100\\
   LWS-005&84$\times$84&200 E&51.8$\mic$/88.4$\mic$&1.24&200-300\\
   LWS-007&84$\times$84&200 N-E&51.8$\mic$/88.4$\mic$&1.15&$\sim$200\\

   \hline

\end{tabular}
\end{center}
\footnotesize
$^{a}$ In respect to the star position.\\

\end{table*}

The mid-IR lines of [SIII] can be used for electron density diagnostics in a
nebula.
In fact, \citet{1984Houck}, \citet{1980Moorwood} and \citet{2002Mizutani} and then
\citet{2009SmithCassiopeaA} computed the [SIII] 18.7 $\mic$/[SIII] 33.5 $\mic$
line ratio dependence on the electron density. They found that this ratio,
temperature-independent, is sensitive to change on the electron density in the
range $100\le n_{e}\le10^{4}\,\rm cm^{-3}$. Hence we have derived the [SIII] 18.7 $\mic$ and 33.5 $\mic$ line fluxes present
in the \emph{ISO} SWS spectrum, positioned 
approximately on the nebula centre (Fig. \ref{fig:slit}). We remind the reader that also in this case background observations are not available. 
The two transitions have measured and dereddened fluxes as in Table
\ref{tab:swslines}. Interstellar absorption is yet important at 18 $\mic$ and
for this reason we consider the dereddened values.

The derived line ratio is F(18.7 $\mic$)/F(33.5 $\mic$)= 0.74$\pm$0.47 which
corresponds to an electron density of $\sim250\,\rm cm^{-3}$.
 The error in this ratio is due to the uncertainty in the measured lines and is
derived using the error propagation for independent measures.  
Similarly, by using the same line pair present in the IRS spectra on the central source (in particular
the line at 33.5 $\mic$ from the LH spectrum and the line at 18.7 $\mic$ from
the SH one), we found a ratio of 1.45 which corresponds to a density of
$\sim10^3\,\rm cm^{-3}$. We took into account the difference in areas covered by
the LH and SH modules (respectively, $11.1\arcsec\times22.3\arcsec$ and $4.7\arcsec\times11.3\arcsec$), rescaling the flux at 33.5 $\mic$ by an approximated factor of
$(4.66)^{-1}$.

We notice that the electron density estimate from the IRS line pair is an order
of magnitude higher than the values found from the SWS lines. This may be due to
the fact that the SWS are not background-subtracted but also to the larger area of the SWS slit in respect to the IRS modules: as a consequence
the central value is averaged with smaller densities far from the centre. Our
estimate is also bigger than those of \citet{2010Esteban}, which found an
electron density of $\sim500\,\rm cm^{-3}$. The reason could be the fact that
they did not correct for extinction, which is yet important at 18.7 $\mic$.

The [OIII] 88.4$\mic$-51.8$\mic$ line pair is a more sensitive indicator of the
electron density, with respect to the [SIII] 33.5$\mic$/[SIII] 18.7$\mic$ line pair
\citep{1984Houck}. 
Using the LWS spectra we have determined [OIII] 88.4$\mic$-51.8$\mic$ line
ratios at six different positions on the nebula.

Derived line ratios and electron density estimates are listed in Table
\ref{tab:OIIIlineratios}. Only in two cases the ratio is just below the low
density limit 0.57 \citep{2002Mizutani}, therefore we provide an upper limit
for the electron density. The low electron density region coincides with the
inner shell (the high-ionized gas, observed also in the radio domain). We want to highlight the fact that electron densities estimated with this method must be considered upper limits, as the ISO spectra were not background-subtracted. However, these electron densities are still consistent with the mean value ($32\,\rm cm^{-3}$) found from the
radio map (which probe the inner shell), which however strongly depends on the geometry 
assumed (Section \ref{sec:em_mass}). We notice that at the stellar position the electron density is higher
$\sim1000\,\rm cm^{-3}$ (from IRS line ratio) and decreases to $\sim250\,\rm
cm^{-3}$ till $ \sim100\,\rm cm^{-3}$ next to the inner and outer shell, except
in the north-east part of the second shell, where it increases again to 
$\sim200\,\rm cm^{-3}$. In fact in this region the [OIII] line is particularly
bright, as discussed in the previous subsection. This strengthens the indication
of high density ($n_{e}>100\,\rm cm^{-3}$) clumps in the vicinity of
G79.29+0.46.

Another pair of lines useful for a diagnostic study is the [SiII] 34.8 $\mic$ to
[SIII] 33.5 $\mic$, that indicates changes in the excitation due to distance
from the star \citep{2007Simpson}, but for these lines our SWS dataset cover
only the central position on nebula. However, the resulting ratio is $\sim0.21$,
which is lower than the diffuse interstellar gas value ($\sim2.5$) and instead
is a typical value in HII region \citep{2007Simpson, 2010Esteban}. We 
notice that [SiII] emission is absent in the background-subtracted spectrum from
IRS taken along the nebula.

We point out that this diagnostics is limited not only by the fact that the ISO spectra are not background-subtracted, but also by the uncertainty of the $A_{V}$.  However, we can exclude much higher value for $A_{V}$ (like the one determined from the IR central object, i.e. 18.6) than 5.6: in fact, we found that $A_{V}=18.6$ leads to electron densities (from the mid-IR lines) of the order of $\sim10^{5}\,\rm cm^{-3}$, not expected in LBV nebulae.
On the contrary, the values found in the spectroscopic analysis give an indication that the visual extinction for the nebula can not be much diverse than the assumed value of 5.6.

\section{Modelling the nebula with Cloudy}
\label{sec:sed}

\begin{figure*}
\begin{center} 
 \includegraphics[width=.8\textwidth]{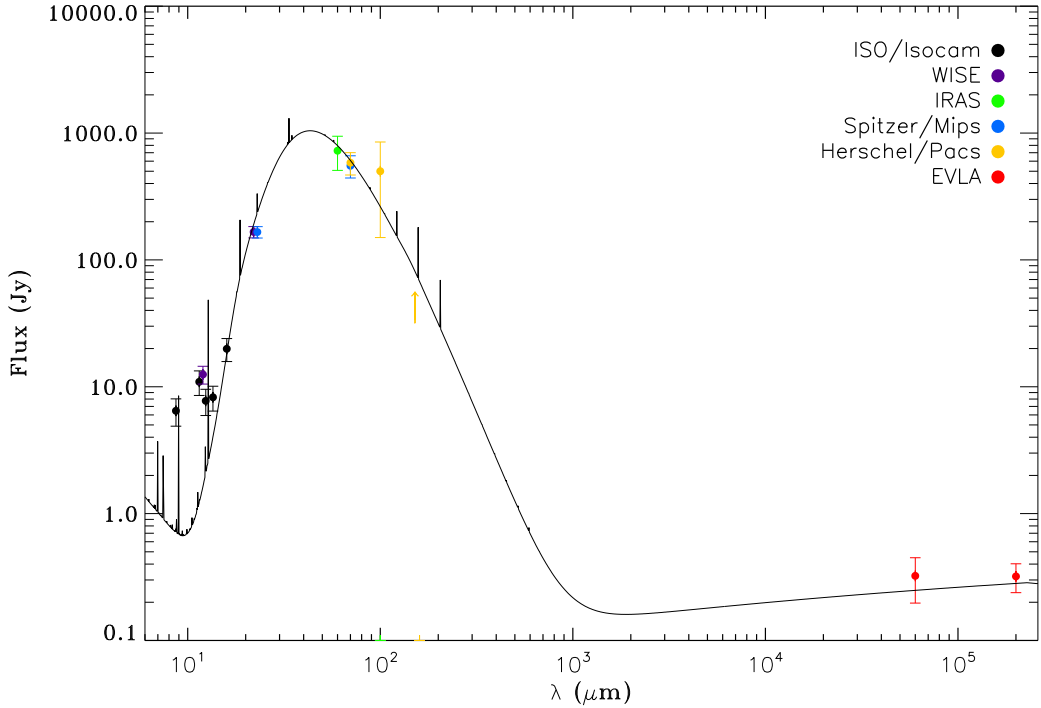}
\caption{Model of the SED and observed flux densities of G79.29+0.46 from
\emph{Spitzer}, \emph{ISO}, \emph{WISE}, \emph{IRAS}, \emph{Herschel} and EVLA observations.}
\label{fig:sed}
\end{center}
\end{figure*}

\begin{table}
\begin{center}
\caption{Model parameters.}
\label{tab:model}

\begin{tabular}{lc}

\hline
  Parameter  & Value \\
 \hline
 T$_{eff}$ (K)& 20400 \\
D (pc) & 1700\\ 
log(r$_{in}$) (cm)& 18.3085 \\
log(r$_{in}$) (pc)& 0.66 \\
M ($\rm M_{\odot}$)&1.53\\
log(L/($\rm L_{\odot}$))&5.4\\
log($\rm n_{H}$) (cm$^{-3}$)&2.13\\
 T$_{e}$ (K)& 5800\\
gas/dust &99\\
     \hline\\
\end{tabular}

\end{center}

\end{table}
Based on the physical parameters derived in our analysis, we have modelled the nebula around G79.29+0.49 with the photo-ionization code CLOUDY version
c10.01, last described by \citet{1998Ferland}. Almost all the assumptions for the model are derived from the observations and this has allowed us to adopt a simple method for the computation of the SED.

The first assumption is that the radiation that illuminates the nebula arises from a black-body with temperature comprises between 18500 K and 20400 K (consistently with a spectral type
B2-B1, as derived in Section \ref{sec:sptype}). The distance adopted is $1.7\,\rm kpc$. The geometry for the nebula is spherical: based on the radio observations it is a shell with inner radius of 0.66 pc. A single shell is assumed. The density is constant within the Str\"{o}mgren radius and then decreases with a law $\propto r^{-3}$. Abundances are chosen as those solar. 
Dust grains are assumed to be composed by both graphite and silicate and described by a power-law distribution \citep{1977Mathis}. By default the grains are resolved into ten size bins. The filling factor is assumed as unity. Cosmic ray are included.

The free parameters are: the stellar luminosity, which has to range between about $10^{5}$ to $10^{6.3}$ $\rm L_{\odot}$, consistently with typical values for LBVs; the grain abundance and the black-body temperature (in the range mentioned above). Therefore, we computed several grids of models and stopped each computation once a nebular mass of 1.53 $\rm
M_{\odot}$ was reached (according to our estimation). Keeping in mind the allowed range for the physical parameters as estimated in this work, we changed them proceeding by iteration in order to explain the observations. In particular, first we set the black-body to 20000 K and varied the hydrogen density (approximated to the electron density, which could range between $\sim30-200\,\rm cm^{-3}$, according to our diagnostics). Starting from the lowest electron density, the first value that fits the radio flux density is then set (in fact, the dust emission resulted to be weakly dependent on the assumed density in respect to the radio). Therefore we varied the grain  abundances  and the stellar luminosity in order to fit respectively the far- and the mid-IR observations. Finally, we adjusted the black-body temperature and the hydrogen density in order to obtain the best model also for the mid-IR spectra.

The final model is shown in Fig. \ref{fig:sed}. The measured flux densities are hence
overlaid. The input and output parameters of the Cloudy model are listed in Table \ref{tab:model}. As shown in the Figure, the model is satisfactory at the longest wavelengths (radio and far-IR, for instance), but at 160 $\mic$ where we provided a lower limit for the flux density (Section \ref{sec:irimages}). The continuum mid-IR emission ($< 15 \mic$) does not describe the data and this might be explained in two  ways.  In the model, small grains are absent because destroyed by the radiation field. In reality they survive because shielded by density clumps (as shown in the maps, the mid-IR emission is clumpy). In support of this there is plenty of observational evidence that very small grains (VSGs) survive within HII regions \citep{2012Paladini, 2011Flagey,2011Paradis} despite theoretical models predict depletion or partial destruction of the grains. Recently also the mid-IR emission of LBV WRA 751 has been explained as due to grains not in thermal equilibrium \citep[e.g.][]{2000Voorsbis, 2013Vamvatira}. Otherwise, and this is more likely, there is lack of hot dust and the observed photometric flux densities are mostly due to the mid-IR lines at wavelengths $< 15 \mic$, as predicted by the Cloudy model. Moreover, the IR continuum in the IRS spectra (Fig. \ref{fig:irs-spectra}) is very little even at the longer wavelengths, suggesting that the observed photometric emission comes from line transitions. The gas-to-dust ratio in the model is $\sim100$, still consistent with the value found in Section \ref{sec:irimages}. 

We have also compared the predicted line ratios with those observed. The considered [OIII] lines are collisionally excited and are sensitive to density. The average observed [OIII] 51.80 $\mic$/[OIII] 88.33 $\mic$ is well predicted by the model, being respectively 0.81 and 0.91, meaning that we set properly the density in the model.  The [FeII] 25.98 $\mic$/[SIII] 33.47 $\mic$ results 0.02 in the model, two orders of magnitude smaller than the observed value 0.3. To understand this discrepancy, we have obtained a grid of models by varying the black-body temperature  and have found that the [FeII] line is controlled by the ionization field and not by abundances. [FeII] 25.98 $\mic$ is very common in PDRs, therefore we suspect that it arises from a layer outer than the ionized gas.
Another interesting finding is that the ratio of [NeII] 12.81 $\mic$/[NeIII] 15.55 $\mic$ observed in the centre of the nebula is 2.2, while the model predicts a higher ratio ($\sim$700), because the photons are not sufficiently energetic to cause [NeIII] transitions. This strongly suggests the presence of collisional excitation to produce both the [NeII] and [NeIII] transitions. 

Line emission of [OI] 63.18 um is not predicted by the model. Therefore the observed lines in the LWS spectra are very probably of interstellar origin.

As a result of the model, the average electron temperature is 5800 K. The main heating process is that photoelectric and the temperature of graphite grains over radius varies from $90\,\rm K$ (at the inner radius) to $57\,\rm K$ (at the outer radius), while for
silicate grains from  $73\,\rm K$ to  $53\,\rm K$. These values are
consistent with our average estimates from the far-IR images (see Table
\ref{tab:tempmass}).

Finally, the derived stellar luminosity (log(L/L$\odot$)=5.4) is consistent with the value provided by different authors \citep[from 5 to 6, e.g.][]{1996Waters,2000Voors} and with that of galactic LBVs \citep{1994HD}. 

The proposed model is based on the assumption that the nebula is a single-shell, while the morphology clearly shows that it consists of multiple shells. However, since the bulk of the determined mass is distributed within the inner shell (at least $98\%$), this simplifying assumption does not affect the model results. To verify this, we have also modelled multiple shells, by considering first the mass contained in the inner shell and giving the transmitted continuum in output as incident radiation field for the second shell. We found a good agreement with the simple model.

The neutral mass, probably distributed in the outer shell, has not been taken into account in the total mass computation. Considering this, the total mass could be higher, as well as the gas-to-dust ratio. A similar model, with same luminosity and effective temperature for the star, still explain the observations up to a mass of $5\,\rm M_{\odot}$ by adjusting properly the gas-to-dust ratio.

\section{Discussion and conclusions}
\label{sec:discus}
We have proposed a multiwavelength analysis of the nebula around the LBV candidate G79.29+0.46 by considering data in a wide spectral range (from the radio to the mid-IR). The aim of this work was to explain all the observed properties under the same hypothesis and to provide a comprehensive picture of the nebular properties.

Summarizing:\\
1) radio observations and the model obtained with the code Cloudy have allowed as to constrain the physical parameters of the star, which has a luminosity of $\sim10^{5.4}\,\rm L_{\odot}$ and an effective temperature of $20.4\times10^3\,\rm K$;\\
2) the star is currently loosing mass through stellar winds (as provided by analysis of the central object spectral index) with a rate of $1.4\times10^{-6}(\frac{v_{\infty}}{110\,\rm
km\,s^{-1}})\ (\ \frac{D}{1.7\,\rm kpc})\ ^{3/2} f^{1/2}\,\rm M_{\odot}\,yr^{-1}$;\\
3) analysis of the IR emission of the central object reveals that the star is surrounded by a close dusty envelope with temperature in the range from $\sim$ 40 to $\sim$ 1200 K; \\
4) at a distance of 0.66 pc there is a detached shell of optical thin ionized gas which emits for free-free encounters. The ionized mass is $\sim1.51\,\rm M_{\odot}$ and the average density is $32\,\rm cm^{-3}$;\\
5) almost co-spatial with the radio emission (Paper 1), there is a component emitting in the mid-IR, which is very likely due to line transitions (HeII, [NeII], [NeIII]) as found in this work. If a component of hot ($T\sim340-550\,\rm K$) thermal dust is present, it has a negligible mass. We exclude the presence of non-thermal dust;\\
6) a component of warm ($T\sim 60-85\,\rm K$) grains is distributed in three concentric shells and has a total mass of $\sim0.02\,\rm M_{\odot}$ (in the case of mixed grains of silicate and graphite);\\
8) no cool dust is evident in the far-IR images;\\
9) the IR spectra show instead the presence of a PDR, very likely distributed outer than the ionized region, but the considered datasets do not allow us to evaluate the contribution of the interstellar material. The electron density in the second shell is yet lower than $\sim 100\,\rm cm^{-3}$, with the exception of the outer region in the north-east, where the density is higher ($\sim200\,\rm
cm^{-3}$). This region corresponds to the radio ``spur-emission", first detected by \citet{1994Higgs}
and present also in our radio maps . The spectral index analysis of
this region gives evidence of the presence of density clumps as well. These clumps could be explained as region modified by the impact of the ejecta with the local environment;\\
10) the gas-to-dust ratio in the nebula is $\approx80 +\rm \frac{M_{neutral [M_{\odot}]}}{0.02}$.

The nebula morphology suggests that the star has lost mass in at least three different episodes. As already reported in Paper 1, two of them occurred about  $5.4 \times 10^{4}\,\rm yr$ and $2.7  \times 10^{4}\,\rm yr$ ago \citep[assuming an expansion velocity of the shell of $\sim30\,\rm km\,s^{-1}$,][]{1996Waters} and now we indicate an intermediate episode of $4.0 \times 10^{4}\,\rm yr$. Considering these epochs, the current mass-loss does not explain the mass contained in the nebula. Assuming a linear expansion, this rate would have produced, in fact, a nebula of mass about few  $10^{-2}\,\rm M_{\odot}$, i.e. two order of magnitude less massive than the value determined from the observations. This implies that a higher mass-loss rate took
place to form such massive nebula and is consistent with changes expected in the rate of mass-loss for LBVs and observed in the case of AG Car.

\citet{1996Waters} reported, by examining IRAS images,
that the nebulae were ejected with a mass-loss rate of $ \sim 5 \times
10^{-4}\,{\rm M_{\odot} yr^{-1}}$. By considering the nebular mass estimated in this work, we suggest
instead a rate of $\sim6.5\times 10^{-5}\,{\rm M_{\odot}
yr^{-1}}$, which is consistent with many galactic LBVs. Mass-loss rates of the order of $ \sim 10^{-5}\,{\rm M_{\odot} yr^{-1}}$ have been derived for AG Car \citep{2009Groh}, the Pistol star and FMM362 \citep{2009Najarro}, G24.73+0.69 \citep{2003Clark}, instead higher mass-loss (greater than $ \sim 10^{-4}\,{\rm M_{\odot} yr^{-1}}$) have been determined for the more luminous $\eta$ Car \citep{2001Hillier}, AFGL 2298 \citep{2009Clark} and G26.47+0.02 \citep{2012Umana}. Lower mass-loss rates ($ \sim 10^{-6}\,{\rm M_{\odot} yr^{-1}}$) were also observed in few cases, e.g. W243 \citep{2009Ritchie}, Wra 17-96 \citep{2002Egan} and HD162685 \citep{2010Umana}.

In the computation of the total nebular mass we have not considered the contribution of the neutral gas. 
Therefore, our estimate of the nebular mass and the average mass-loss rate must be considered lower limits. However, the discrepancy with \citet{1996Waters} is probably mostly due to a difference in the ionised mass determined in previous works with respect to our estimate. In fact, the mass determined in this work is at least four times smaller than the mass computed by \citet{1994Higgs}. A massive nebula, such that proposed previously, would be comparable in mass to the Homunculus nebula around $\eta$ Car, despite G79.29+0.46 is less luminous, and consequently less massive, than $\eta$ Car. While the difference between the high nebular mass around G79.29+0.46 and its initial mass would have fascinating implications for massive stellar evolution, the ejecta mass estimated in this work appears to be in agreement with a 'low-luminosity' massive star.

G79.29+0.46 is still considered a candidate LBV because it does not match the variability criteria. However, indirect evidence for changes in the mass-loss rates have been pointed out. Indeed it could be a LBV in the quiescent state and which suffered at least three events of higher mass-loss over the last $5.4\times10^{4}$ yr.

\section*{Acknowledgments}

We thank the anonymous referee for useful comments that helped to improve the manuscript. This paper is part of the PhD thesis of C.A., who is very thankful to her tutors for supervising her PhD work. C.A. also thanks the \emph{Spitzer} Science Center/Infrared Processing and Analysis Center staff at Caltech for the support provided during her visit to complete this project, and address a special thanks to Roberta Paladini for the invaluable suggestions received  during her PhD program. 

This work is based on observations performed at the National Radio Astronomy Observatory, a facility of the National Science Foundation operated under cooperative agreement by Associated Universities, Inc. \emph{Herschel} is an ESA space observatory with science instruments provided by European-led Principal Investigator consortia and with important participation from NASA. This research has also made use of the NASA/IPAC Infrared Science Archive, which is operated by
the Jet Propulsion Laboratory, California Institute of Technology, under
contract with the National Aeronautics and Space Administration. \emph{ISO} was an ESA project with
instruments funded by ESA Member States (especially the PI countries: France,
Germany, the Netherlands and the United Kingdom) and with the participation of
ISAS and NASA. Finally, this publication makes use of data products from the Two Micron All Sky Survey, which is a joint project of the University of Massachusetts and the Infrared Processing and Analysis Center/California Institute of Technology, funded by the National Aeronautics and Space Administration and the National Science Foundation.

\bsp

\clearpage

\appendix

\section{The IR spectra}

\begin{figure}
\begin{center} 
\begin{minipage}{.9\linewidth}
    \includegraphics[width=\textwidth]{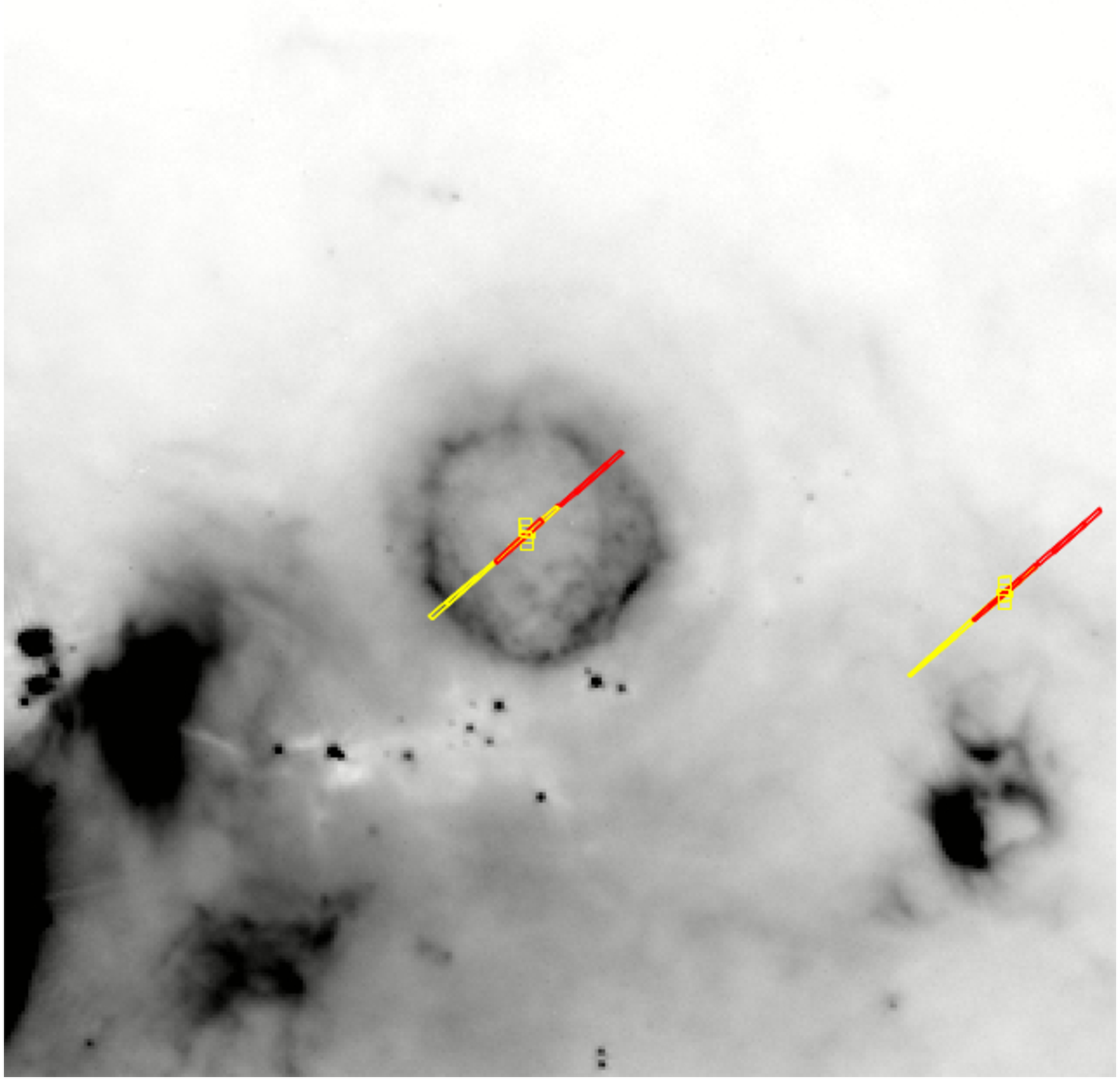}
   \caption{MIPS/\emph{Spitzer} 24 $\mic$ image of G79.29+0.46 and positions of
the IRS-SH and LH/\emph{Spitzer} (centre) and IRS-SL/\emph{Spitzer} (nebula)
slits.}
\label{fig:slitirs}
 \end{minipage}
\end{center}
\end{figure}

\begin{figure}
\centering
\begin{minipage}{.9\linewidth}
\vspace{1cm}    
\includegraphics[width=\textwidth]{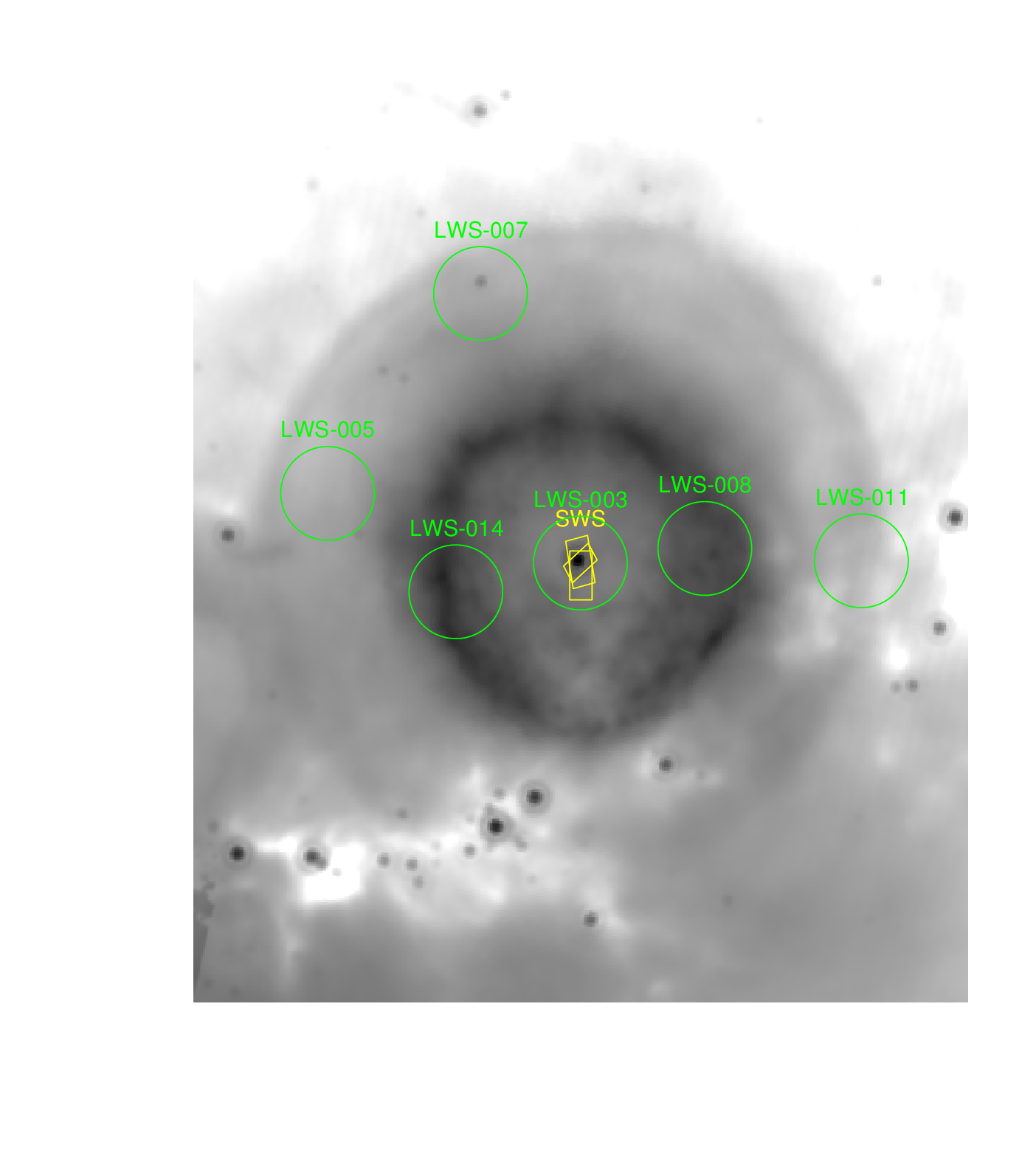}
   \caption{MIPS/\emph{Spitzer} 24 $\mic$ image of G79.29+0.46 and positions of
the LWS/\emph{ISO} (green) and SWS/\emph{ISO} (yellow) slits.}
\label{fig:slit}
 \end{minipage}
\end{figure}

\begin{figure}
\centering
 \begin{minipage}{1\linewidth}
 \includegraphics[width=\textwidth]{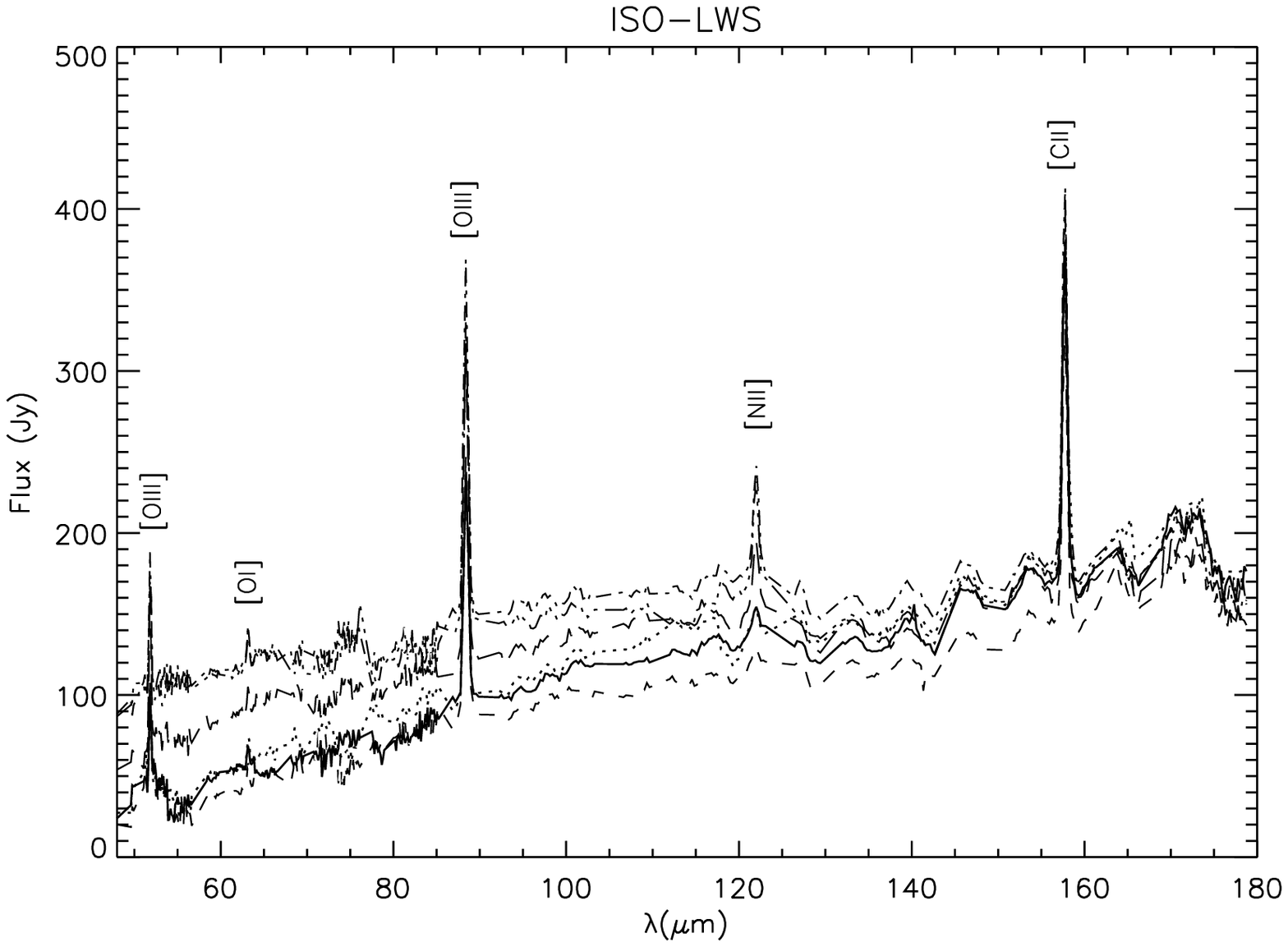}
  \caption{LWS spectra taken at six different position on the nebula (three on the first shell and three on the second shell)}. 
\label{fig:lws-spectra}
 \end{minipage}

\end{figure}

\begin{figure}
\centering
 \begin{minipage}{1\linewidth}
 \includegraphics[width=\textwidth]{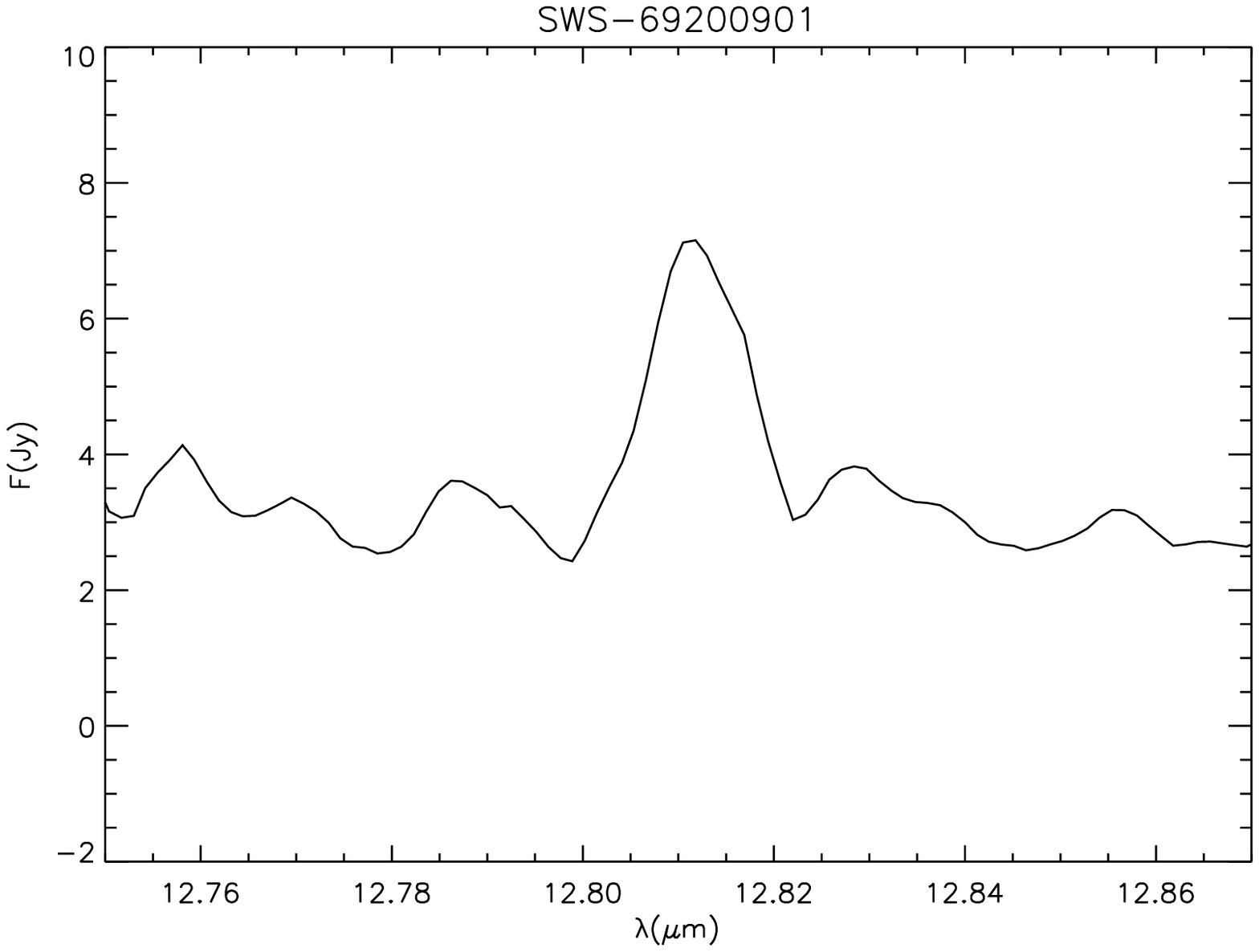}
  \vspace{4pt}
 \includegraphics[width=\textwidth]{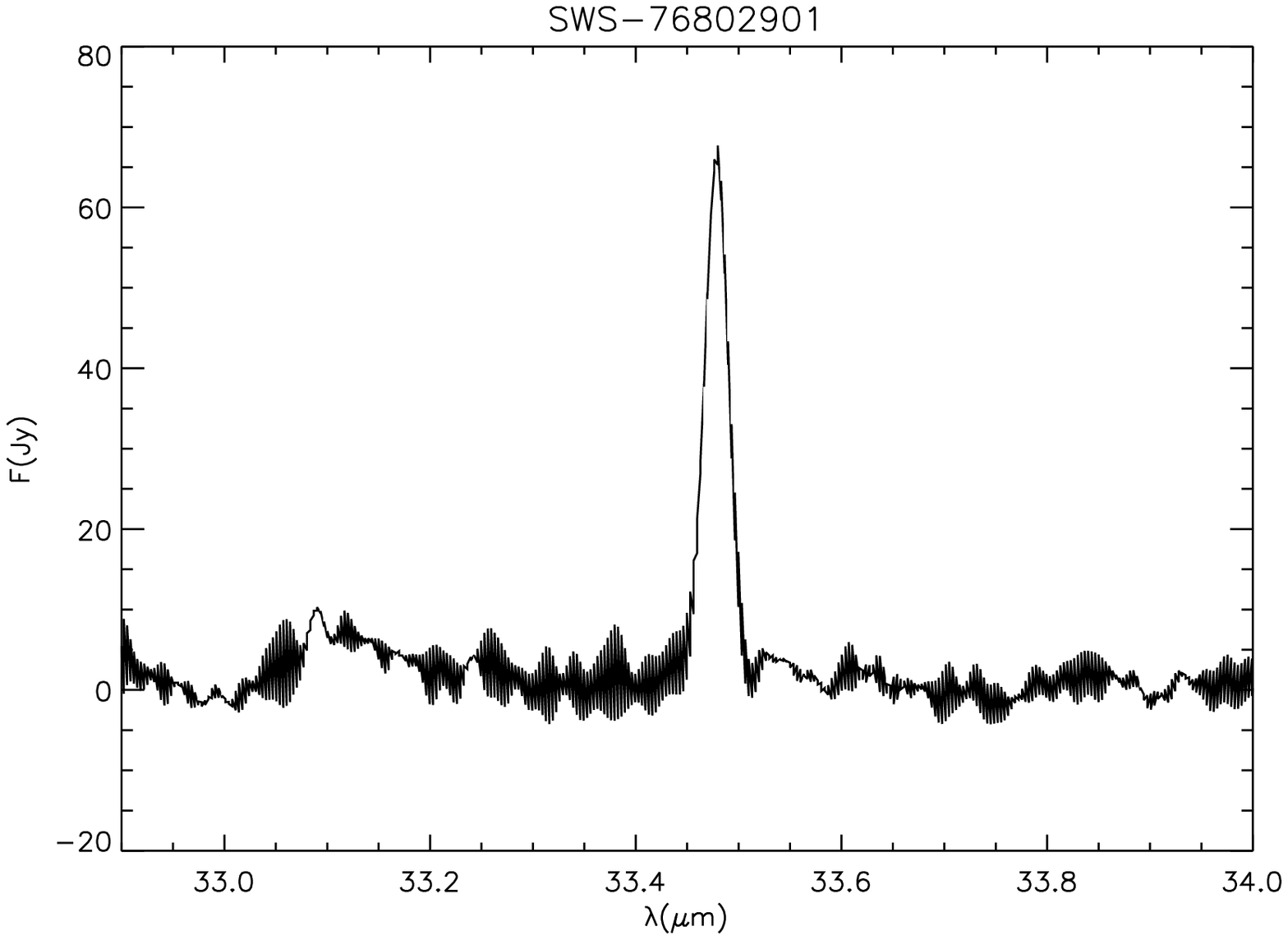}
  
 \includegraphics[width=\textwidth]{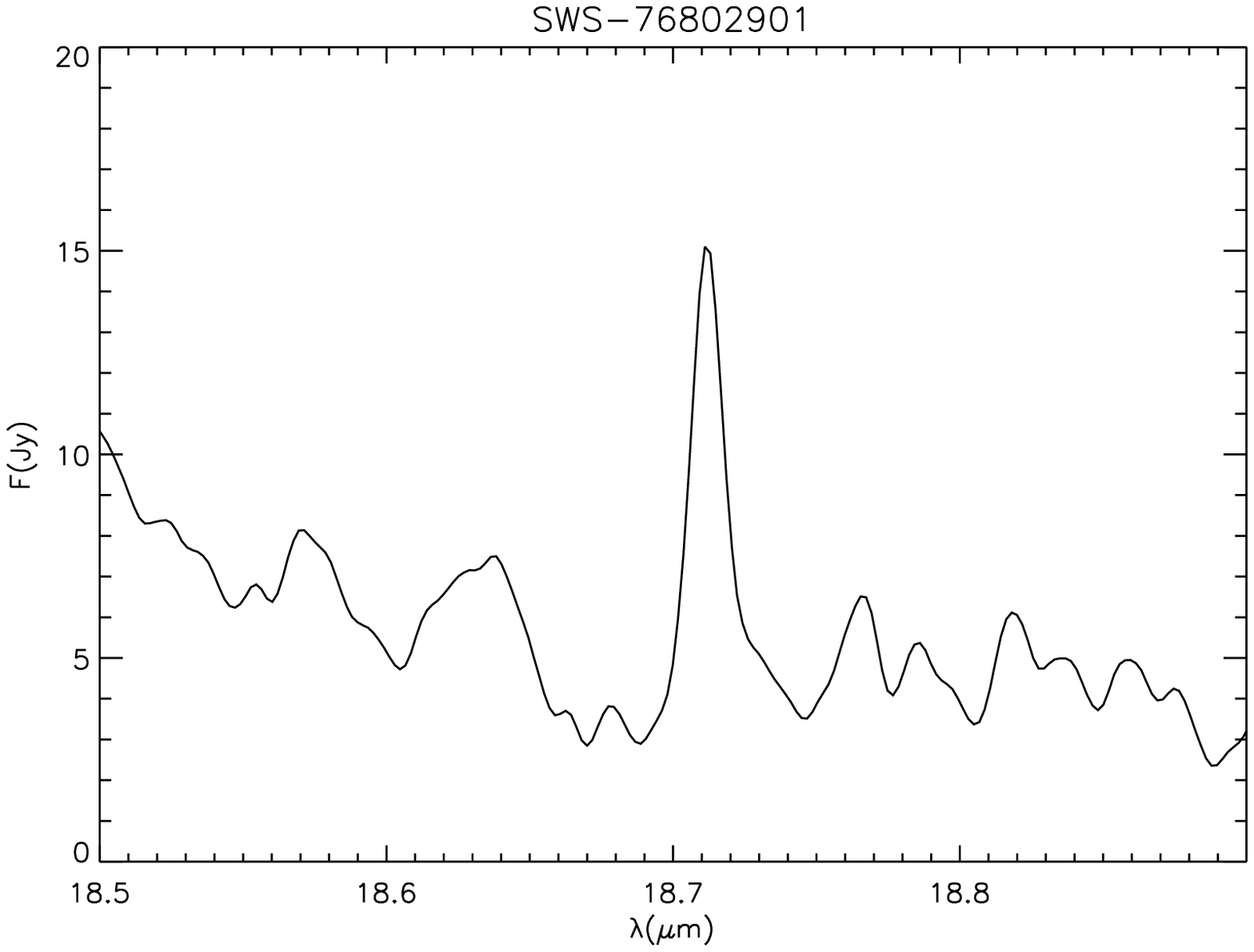}

 \end{minipage}
 \caption{\emph{ISO}-SWS spectra taken on the star position as shown in Fig.
\ref{fig:slit}. We do not show the [SiII] 34.81 $\mic$ line because affected by fringing.}
 \label{fig:sws-spectra}
\end{figure}

\begin{figure}
\centering
 \begin{minipage}{1\linewidth}

\includegraphics[width=\linewidth]{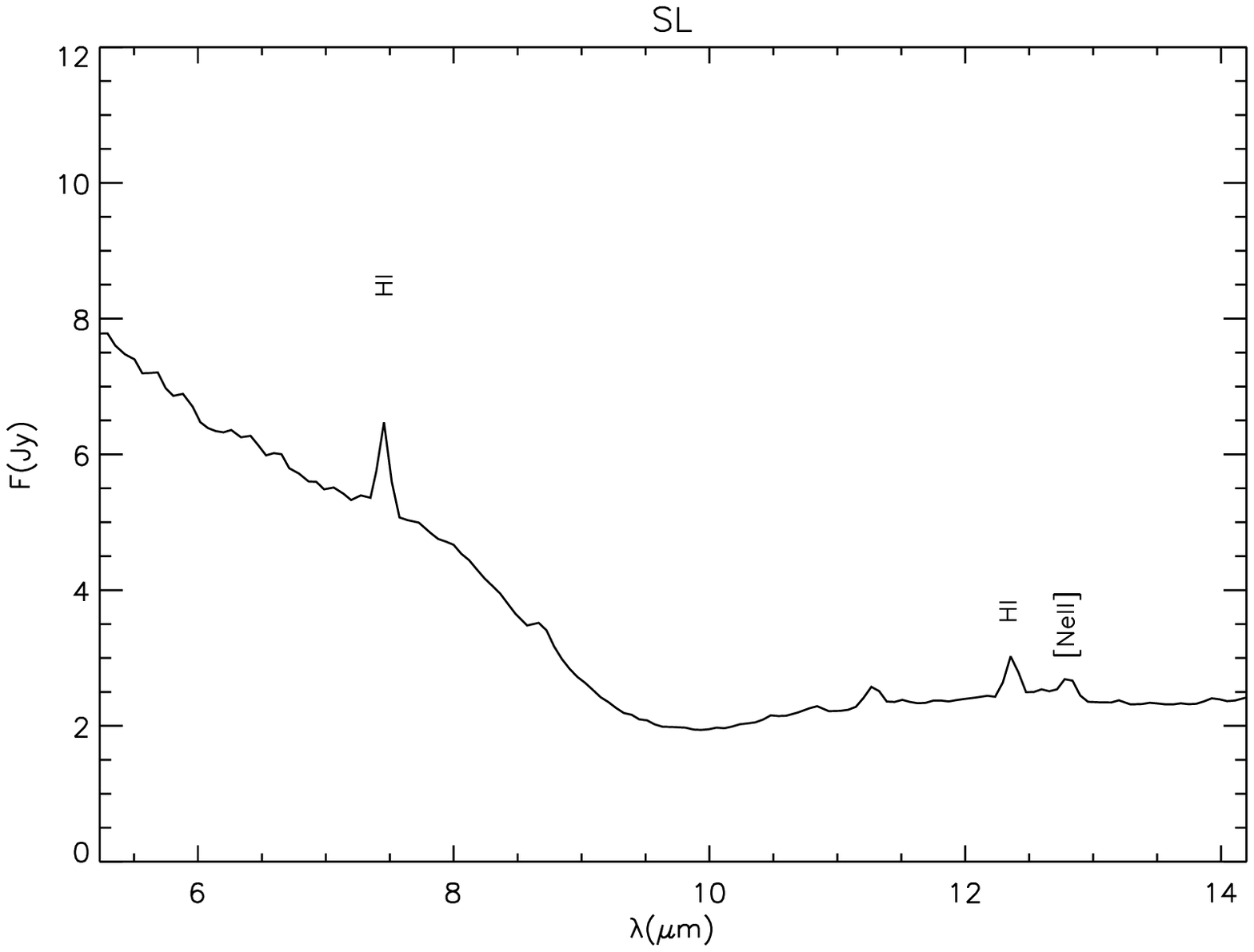}
 \includegraphics[width=\linewidth]{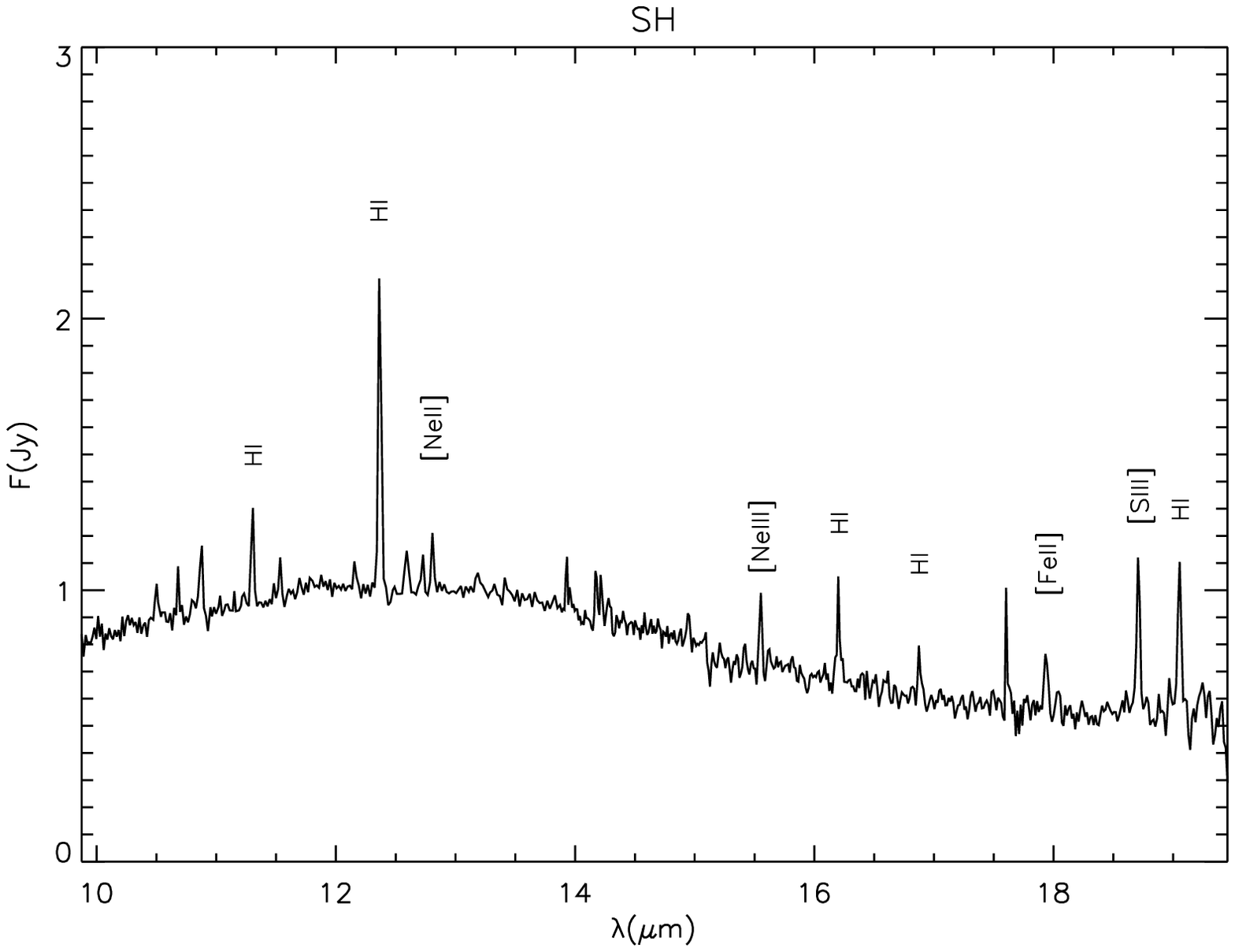}
 \includegraphics[width=\linewidth]{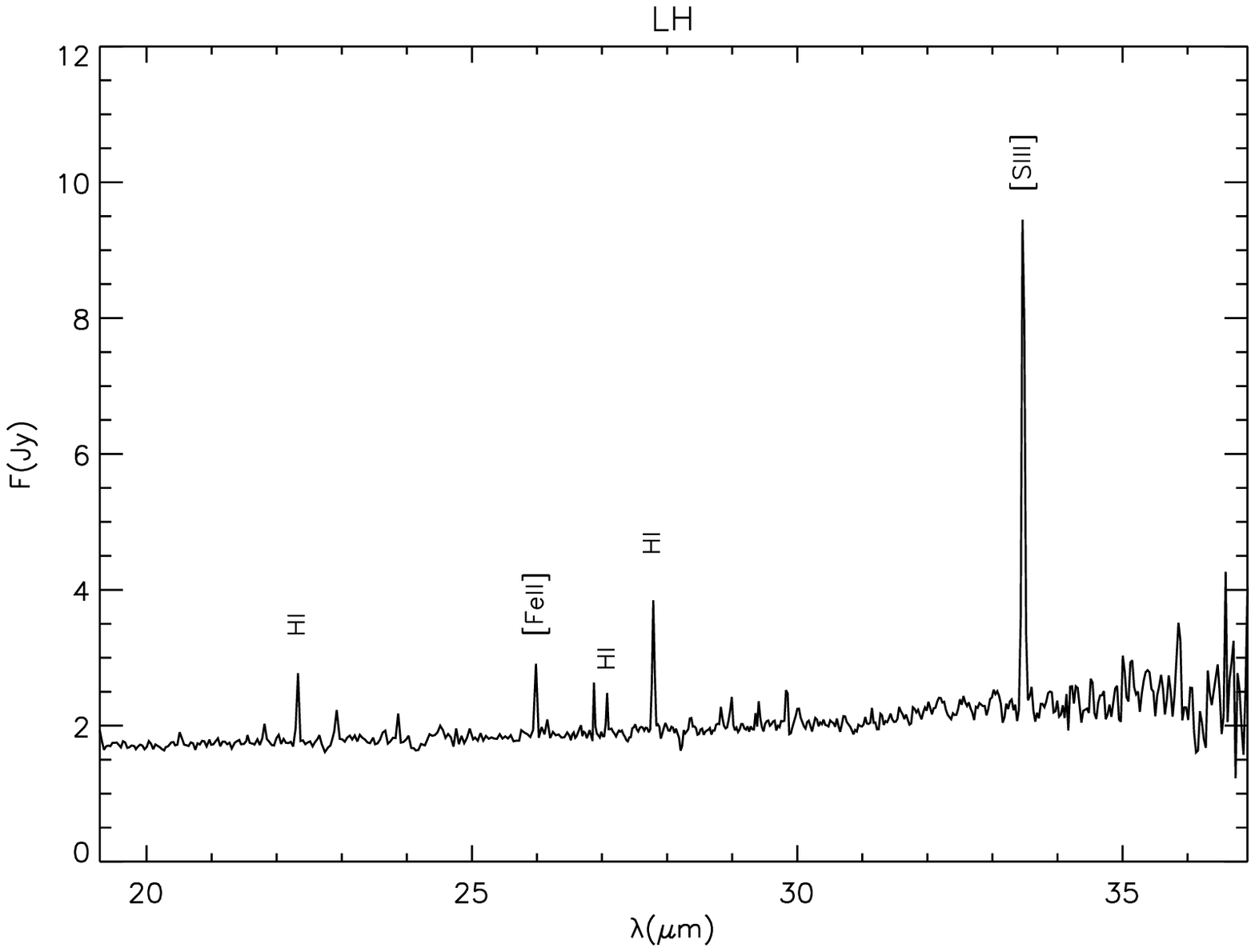}
  \caption{IRS SL (CH0), SH (CH1), LH (CH3) spectra after the background
subtraction. Data were taken on the star position (SL and SH) and on the nebula (LH) as shown in Fig. \ref{fig:slitirs}. }
 \label{fig:irs-spectra}
  \end{minipage}

\end{figure}

Fig. \ref{fig:slitirs} and \ref{fig:slit} show the position of the \emph{Spitzer} and ISO slits, respectively, superimposed on the 24 $\mic$ \emph{Spitzer} image. In Fig. \ref{fig:lws-spectra} all the LWS spectra, taken at different positions on nebula,  are shown. Fig. \ref{fig:sws-spectra} contains the most significant emission lines detected by the SWS instrument positioned at the nebula centre. The IRS background subtracted spectra are shown in Fig. \ref{fig:irs-spectra}. Finally, the Tables list the line fluxes measurements before and after the extinction correction.    
\clearpage

\begin{table*}
\begin{center}
\begin{small}
\caption{SWS line fluxes.}
\label{tab:swslines}

\begin{tabular}{llccllc}
   \hline
   $\lambda_{cent}$&$\lambda_{lab}$& Element & Transition &
F$_{observed}$&F$_{dereddened}$&\# \\
   $_{(\mic)}$ & $_{(\mic)}$& & & ($_{10^{-19}\rm W\, cm^{-2}}$)&(
$_{10^{-19}\rm W\, cm^{-2}}$)&\\
   \hline

12.8121&12.8135&[NeII]&2P$\frac{1}{2}$-2P$\frac{3}{2}
$&1.05$\pm$0.02&3.2$\pm$0.4&sws69200901\\  
   18.7123&18.7130&[SIII]&3P2-3P1&1.5$\pm$0.1&4.6$\pm$0.6&sws76802901\\ 
   33.4780&33.4810&[SIII]&3P1-3P0&5.4$\pm$0.1&6.2$\pm$0.2&sws76802901\\
  
34.8128&34.8152&[SiII]&2P$\frac{3}{2}$-2P$\frac{1}{2}
$&1.3$\pm$0.3&1.3$\pm$0.1&sws51705105\\
   \hline

\end{tabular}
\end{small}
\end{center}
\end{table*}

\begin{table*}
\begin{center}
\begin{small}
\caption{Derived line fluxes from LWS spectra, obtained at six different
position of the slit on the nebula.}
\label{tab:lwslines}

\begin{tabular}{llccllc}
   \hline
   $\lambda_{cent}$&$\lambda_{lab}$& Element & Transition &
F$_{observed}$&F$_{dereddened}$&\# \\
   $_{(\mic)}$ & $_{(\mic)}$& & & ($_{10^{-19}\rm W\, cm^{-2}}$)&($_{10^{-19}\rm
W\, cm^{-2}}$)&\\
   \hline
   
   51.8333&51.8145&[OIII]&3P2-3P1&20.1$\pm$3.6&28.2$\pm$4.8&lws35801011\\  
   51.8191&&&&30.8$\pm$4.4&55$\pm$5&lws35801007\\ 
   51.7381&&&&38.7$\pm$5.5&46$\pm$6&lws35801005\\
   51.8234&&&&29.6$\pm$2.4&32.6$\pm$3.7&lws35801014\\
   51.8471&&&&26$\pm$2&28.5$\pm$3.1&lws35801008\\
   51.8221&&&&34.4$\pm$3.4&38.8$\pm$3.4&lws35801003\\
   \hline
   63.1785&63.1837&[OI]&(3P1-3P2)&4.6$\pm$0.8&4.6$\pm$1.4&lws35801011\\
   63.1785&&&&3.9$\pm$0.3&3.3$\pm$1.8&lws35801007\\
   63.2027&&&&3.7$\pm$0.8&4.1$\pm$1.1&lws35801005\\
   63.1774&&&&5.1$\pm$1.2&5.3$\pm$1.1&lws35801014\\
   63.1734&&&&5.3$\pm$1.8&5.4$\pm$2.0&lws35801008\\
   63.1870&&&&3.3$\pm$0.3&4.7$\pm$2.3&lws35801003\\
   \hline
   88.4118&88.3560&[OIII]&(3P1-3P0)&40.8$\pm$2.7&42$\pm$3&lws35801011\\
   88.3803&&&&43.05$\pm$1.46&47.7$\pm$5.7&lws35801007\\
   88.3960&&&&35$\pm$2&37.2$\pm$3.6&lws35801005\\
   88.3734&&&&59.4$\pm$5.1&61.5$\pm$6.1&lws35801014\\
   88.4099&&&&51.2$\pm$1.8&51.7$\pm$4.2&lws35801008\\
   88.3817&&&&52.2$\pm$1.1&53$\pm$3&lws35801003\\
   \hline
   122.1056&121.898&[NII]&(3P2-3P1)&8$\pm$1&8.4$\pm$2.3&lws35801011\\   
   122.0106&&&&3.6$\pm$1.2&4.4$\pm$1.5&lws35801007\\
   121.6556&&&&4.7$\pm$1.7&5.1$\pm$1.8&lws35801005\\
   121.9799&&&&12.5$\pm$1.5&12.9$\pm$2.2&lws35801014\\
   121.9589&&&&14.2$\pm$2.5&15$\pm$2&lws35801008\\
   121.9886&&&&8.79$\pm$0.76&10.0$\pm$1.6&lws35801003\\
   \hline
  
157.7338&157.741&[CII]&2P$\frac{3}{2}$-2P$\frac{1}{2}
$&17.6$\pm$0.1&18.1$\pm$2.2&lws35801011\\
   157.7237&&&&15.01$\pm$0.35&15.2$\pm$1.7&lws35801007\\
   157.7191&&&&18.0$\pm$1.5&18.6$\pm$1.6&lws35801005\\
   157.7185&&&&20.2$\pm$0.6&20.7$\pm$1.6&lws35801014\\
   157.7289&&&&18.7$\pm$1.2&19.9$\pm$2.8&lws35801008\\
   157.7207&&&&17.2$\pm$0.4&17.5$\pm$1.7&lws35801003\\

   \hline

\end{tabular}
\end{small}

\end{center}

\end{table*}

\begin{table*}
\begin{center}
\begin{small}
\caption{Line fluxes derived from the IRS/SL spectrum, after the background
subtraction.}
\label{tab:ch0lines}

\begin{tabular}{llccll}
   \hline
   $\lambda_{lab}$&$\lambda_{cent}$& Element & Transition &
F$_{observed}$&F$_{dereddened}$\\
   $_{(\mic)}$ & $_{(\mic)}$& & & ($_{10^{-21}\rm W\, cm^{-2}}$)&($_{10^{-21}\rm
W\, cm^{-2}}$)\\
   \hline
   7.4568&7.4526&HeII&\footnotesize{
12-10}&962$\pm$160&\footnotesize{(2.5$\pm$1.4)$\times10^{3}$}\\ 
   12.2485&12.3635&HeII&\footnotesize{25-17
}&118.9$\pm$29.6&485.7$\pm$23.3\\  
   12.8135&12.8070&[NeII]&\footnotesize{2P$\frac{1}{2}$-2P$\frac{3}{2}$}&82.9$
\pm $10.7&135.3$\pm$105.8\\
   
   \hline

\end{tabular}
\end{small}

\end{center}
\end{table*}

\begin{table*}
\begin{center}
\caption{Line fluxes derived from the IRS/SH spectrum, after the background
subtraction.}
\label{tab:ch1lines}

\begin{tabular}{llccll}
   \hline
   $\lambda_{lab}$&$\lambda_{cent}$& Element & Transition &
F$_{observed}$&F$_{dereddened}$\\
   $_{(\mic)}$ & $_{(\mic)}$& & & ($_{10^{-21}\rm W\, cm^{-2}}$)&(
$_{10^{-21}\rm W\, cm^{-2}}$)\\
   \hline

   11.3026&11.3025&HI (HeII)&23-16 (18-14)&27.8$\pm$5.6&165.3$\pm$39.9\\  
   12.3669&12.3677&HI (HeII)&14-12 (7-6)&65.4$ \pm $14.0&275.5$\pm$17.7\\
   12.8135&12.8086&[NeII]&2P1/2-2P3/2&9.9$ \pm $1.5&40.0$\pm$13.0\\
   15.5551&15.5511&[NeIII]&3P1-3P2&11.6$ \pm $3.4&17.9$\pm$3.2\\
   16.2025&16.2009&HI&10-8&10.1$ \pm $2.2&24.0$\pm$3.6\\
   16.8737&16.8760&HI&12-9&4.3$ \pm $1.1&13.2$\pm$3.7\\
   17.9359&17.9358&[FeII]&a4F7/2-a4F9/2&8.1$ \pm $2.2&24.9$\pm$7.0\\
   18.7130&18.7085&[SIII]&3P2-3P1&14.8$ \pm $1.8&45.6$\pm$6.1\\
   19.0541&19.0527&HI&8-7&15.5$ \pm $3.6&46.9$\pm$10.1\\

   \hline

\end{tabular}

\end{center}
\end{table*}

\begin{table*}
\begin{center}
\begin{small}
\caption{Line fluxes derived from the IRS/LH spectrum, after the background
subtraction.}
\label{tab:ch3lines}

\begin{tabular}{llccll}
   \hline
   $\lambda_{lab}$&$\lambda_{cent}$& Element & Transition &
F$_{observed}$&F$_{dereddened}$\\
   $_{(\mic)}$ & $_{(\mic)}$& & & ($_{10^{-21}\rm W\, cm^{-2}}$)&(
$_{10^{-21}\rm W\, cm^{-2}}$)\\
   \hline
  
   22.3316&22.3191&HI (HeII)&13-10 (22-18)&24.5$\pm$7.7&61.9$\pm$14.2\\  
   25.9883&25.9758&[FeII]&a6D7/2-a6D9/2&24.1$\pm$2.9&42.4$\pm$15.2\\
   27.0880&27.0756&HI&24-14&7.2$\pm$1.5&13.3$\pm$5.3\\  
   27.7921&27.7887&HI&9-8&30.7$\pm$3.3&54.6$\pm$7.1\\    
   33.4810&33.4729&[SIII]&3P1-3P0&109.9$\pm$6.3&146.5$\pm$5.4\\
   \hline

\end{tabular}
\end{small}

\end{center}
\end{table*}

\label{lastpage}

\end{document}